\documentclass[12pt]{article}
\usepackage[dvips]{color}
\usepackage{amsmath}
\usepackage{graphicx}
\usepackage{subfigure}
\usepackage{epsfig}
\usepackage{amsmath}
\usepackage{cite}
\usepackage{color}
\usepackage{subfigure}
\usepackage{multirow}

\begin{document}
\begin{center}
\Large{\bf Phase Transition Dynamics of Black Holes Influenced by Kaniadakis and Barrow Statistics}\\
 \small \vspace{1cm}
 {\bf Jafar Sadeghi $^{\star}$\footnote {Email:~~~pouriya@ipm.ir}}, \quad
 {\bf Mohammad Ali S. Afshar $^{\star}$\footnote {Email:~~~m.a.s.afshar@gmail.com}}\quad
 {\bf Mohammad Reza Alipour $^{\star}$\footnote {Email:~~~mr.alipour@stu.umz.ac.ir}}\quad
 \\\vspace{0.2cm}
 {\bf Saeed Noori Gashti$^{\dag,\star}$\footnote {Email:~~~saeed.noorigashti@stu.umz.ac.ir}}, \quad\\
\vspace{0.9cm}$^{\star}${Department of Physics, Faculty of Basic
Sciences,\\
University of Mazandaran
P. O. Box 47416-95447, Babolsar, Iran}\\
\vspace{0.5cm}$^{\dag}${School of Physics, Damghan University, P. O. Box 3671641167, Damghan, Iran}
\small \vspace{1cm}
\end{center}
\begin{abstract}
In this study, we investigate the dynamics and frame-by-frame phase transition of the first order in black hole thermodynamics. For our analysis, we will utilize the Kramers escape rate. Our focus is on charged anti-de Sitter (AdS) black holes influenced by Kaniadakis and Barrow statistics. The selection of these black holes aims to examine the effects of entropy variation on the dynamics of phase transition, and to demonstrate that the Kramers escape rate, as an efficient tool, can effectively represent the dynamic transition from a small to a large black hole within the domain of first-order phase transitions. It is noteworthy that while the transition from small to large black holes should ostensibly dominate the entire process, our results indicate that the escape rate undergoes changes as it passes through the midpoint of the phase transition, leading to a reverse escape phenomenon. The findings suggest that the dynamic phase transition in charged AdS black holes affected by entropy change bears a significant resemblance to the outcomes of models influenced by Bekenstein-Hawking entropy\cite{23}. This similarity in results could serve as an additional motivation to further explore the potential capabilities of Kaniadakis and Barrow statistics in related cosmological fields. These capabilities could enhance our understanding of other cosmological properties.
\tableofcontents
\end{abstract}
\section{Introduction}
Black holes, those cosmic enigmas born from gravitational collapse, serve as fascinating laboratories for probing quantum phenomena. Among the key aspects under scrutiny is the thermal entropy of black holes, a quantity directly tied to their event horizon area. Over decades of development, black hole thermodynamics has acquired a relatively complete theoretical framework\cite{1}. While many questions remain unanswered, it provides a valuable window into the mysteries of quantum gravity. The thermodynamic phase transition and the behavioral similarity of the dimensional transition of black holes in the anti-de Sitter (AdS) space-time to this process have greatly fascinated researchers\cite{1000,1001,1002,1003,1004,1005,1006,1007,1008,1009}. These transitions intricately connect statistical mechanics, quantum mechanics, and general relativity. Notably, the Hawking-Page phase transition occurs between thermal radiation and the large AdS black hole\cite{2}. This transition has been likened to the confinement/deconfinement phase transition in gauge fields\cite{a}. Black hole thermodynamics is studied in various structures and spaces, including the extended phase space\cite{3,4,5,6}. Drawing an analogy with van der Waals fluids, researchers speculate that theoretical black holes possess a hidden microstructure\cite{7,8,9}. Recent works explore the physical interpretation of black hole thermodynamics using the AdS/conformal field theory correspondence\cite{10,11,12,12'}. This correspondence sheds light on relevant concepts, bridging the gap between field theory and black hole phenomena.\\\\
Existing research primarily centers on analyzing the criticality and type of thermodynamic phase transitions in black holes. However, these studies often neglect detailed descriptions of the actual phase transition process. Recently, researchers have introduced the concept of the free energy landscape to explore related evolutionary processes during black hole phase transitions within the framework of non-equilibrium statistical physics. By calculating the mean first passage time, they have made initial investigations into the kinetics of these transitions\cite{13,14,15,16,17,18,19}. Given that stochastic fluctuations influence some thermodynamic processes, applying methods from non-equilibrium statistical physics can yield crucial insights into their occurrence.\\
Building upon the aforementioned statements, we have embarked on a novel examination of the first-order phase transition dynamics of black holes, grounded in a stochastic framework and employing the Kramers escape rate. Our previous studies on black holes influenced by Bekenstein-Hawking entropy revealed that the Kramers escape rate adeptly represents the dynamics and the probabilistic trend of the transition process from a small to a large black hole within the phase transition domain. Additionally, our prior results indicated the emergence of a reverse trend midway through the phase transition, which progressively evolved\cite{23}. This spurred us to question whether structures undergoing entropy changes would witness alterations in their dynamic phase transition process.
This scenario could culminate in two predictions: on one hand, if the phase transition is subject to change, investigating this shift could be intriguing and open new avenues in the study of phase transitions. On the other hand, if the findings of these studies are consistent and uniform with previous results, it could serve as evidence advocating for the increased utilization of the latent capacities of altered entropies in gravitational and cosmological research, an area that has hitherto received scant attention.
To this end, we specifically scrutinize the phase transition from small to large black holes in AdS black holes, considering the Kaniadakis and Barrow statistics. These transitions resemble the gas-liquid phase transition observed in van der Waals fluids\cite{13,14,15,16,17,18,19}. Our focus centers on understanding the first-order phase transition, which occurs in charged AdS black holes. In equilibrium statistical physics, Gibbs free energy behavior allows us to determine the system's preferred state as temperature changes. The intersection of a swallowtail structure in Gibbs free energy serves as the turning point indicating the favored state\cite{13,14,15,16,17,18,19}. To analyze this, we employ Kramer's escape rate method, which describes the Brownian motion of particles in an external field and thermal potential\cite{20,21,22,23}.\\\\
The concept of thermodynamic entropy plays a crucial role in black hole (BH) physics. According to the holographic principle, BHs can store information at their event horizons, analogous to holograms\cite{24,25}. Researchers have explored extensions of Boltzmann-Gibbs statistics, driven by both gravitational considerations (such as Tsallis and Barrow entropies) and information theory (including Rényi and Sharma-Mittal entropies). These models have undergone testing in cosmology and quantum physics. Recently, Kaniadakis proposed a non-extensive generalization inspired by the symmetries of the relativistic Lorentz group\cite{26,27,28,29,30,31,32,33,34}.
Based on the concepts outlined above, we organize the article as follows.\\ In Section 2, we provide a brief explanation of Kramer's escape rate in relation to the Gibbs free energy and thermal potential. Moving on to Section 3, we introduce the AdS black hole with Kaniadakis statistics and apply the concepts discussed in Section 2 to it. In Section 4, we repeat a similar process, this time focusing on Barrow statistics. Finally, in Section 5, we present the complete results of our work.

\section{Phase Transition Dynamics}
Traditional static methods in phase transition studies, such as the swallowtail bifurcation diagram, provide valuable insights into the thermodynamics of black holes. However, they often overlook dynamic aspects and the temporal sequence of events. Our research focuses on the Kramers escape rate, which offers a more dynamic approach to phase transitions.
In a recent study by \cite{23}, the free energy landscapes for black holes were examined under the influence of 'dark' and 'stringy+dark' structures. The goal was to assess how additional parameters impact escape rates and the dynamics of the first-order phase transition from small to large black holes.
Our analysis considers the escape rate as a function of the black hole radius, revealing interesting variations. On one hand, the escape rate aligns with our assumption: it starts from zero, increases to a maximum point, and then decreases back to zero as reactive structures become active during the phase transition. However, a critical point arises where the direct process (escape rate from small to large black holes) intersects with the reverse process (large to small black holes). This point, seemingly improbable at the onset of the transition, gains significance as the process unfolds. It indicates a region where the escape rate from larger black holes to smaller ones dominates.
Interestingly, the predominance of the reverse process increasing as we approach the end of the transition is a natural reaction of the black hole against the 'phase change.' Maybe this reaction aims to prevent uncontrolled radial growth that could destabilize the black hole. Our investigation, inspired by \cite{23}, focuses on phase transitions in anti-de Sitter (AdS) black holes influenced by Kaniadakis and Barrow statistics. Our work addresses gaps in stochastic process analysis related to phase transition rates and suggests potential applications of Kaniadakis and Barrow statistics in cosmological contexts. Finally, we compare our results with existing literature, including \cite{23}.
\subsection{Kramer's escape rate}
The study of Brownian particle escape under thermal fluctuations from a stable or metastable potential, known as the Kramers problem, is a captivating topic that has attracted attention across various branches of physics. This subject, along with the diverse results it has produced, has led to the formulation of the Kramers escape rate\cite{23}.
our aim is to model the phase transition behavior of black holes, the over-damping mode appears to be the most suitable option\cite{23}. This is because, in this mode, the combined effects are strengthened, making the system unstable and increasing the likelihood of the particle escaping the potential well. Over time, we can expect the particle to have crossed the barrier.
The escape rate is closely related to the inverse of the average time required for the first passage of the local maximum potential, known as the "mean first passage time." According to the referenced articles, the Kramers escape rate can be expressed as follows\cite{23}:
\begin{equation}\label{(1)}
r_{k}=\frac{\mathrm{D} \sqrt{{| \frac{d^{2}}{d x^{2}}U( x_{max}) \! } { \frac{d^{2}}{d x^{2}}U( x_{min}) \! |}}\, {\mathrm e}^{-\frac{U \left(x_{\max}\right)-U \left(x_{\min}\right)}{k_{B} T}}}{2 \pi  k_{B} T},
\end{equation}
where, the $x_{min}$ is a local minimum, local maximum located at $x_{max}$, $U(x)$ is a potential field, $D $ is the diffusion coefficient, $k_B$ is the Boltzmann constant and $T$ is the temperature. When the system reaches thermal equilibrium, the diffusion coefficient \(D\) can be considered constant. With some simplification, the relationship can be rewritten\cite{23},
\begin{equation}\label{(2)}
r_{k}=\frac{ \sqrt{{| \frac{d^{2}}{d x^{2}}U( x_{max}) \! } { \frac{d^{2}}{d x^{2}}U( x_{min}) \! |}}\, {\mathrm e}^{-\frac{U \left(x_{\max}\right)-U \left(x_{\min}\right)}{D}}}{ 2 \pi}.
\end{equation}
\subsection{Gibbs free energy landscape $\&$ thermal potential}
The Gibbs free energy function, known for its dependence on enthalpy and temperature, is a crucial tool for examining phase transitions within black holes, especially under constant pressure conditions. The swallowtail behavior, depicted graphically as a function of temperature, is a common method for exploring these transitions. In the context of free energy, the Gibbs free energy is a continuous function closely linked to the system's order parameter, providing a solid framework for analyzing phase behavior\cite{23},
\begin{equation}\label{(3)}
\begin{split}
G=H-T_{H}S,\\
G_L =M-T S.
\end{split}
\end{equation}
The difference between \( G \) and \( G_{L} \) is reflected in the value of temperature. In the first equation, \( T_H \) represents the Hawking temperature, whereas in the second equation, \( T \) is a temperature that, although applicable to thermodynamic equations, does not necessarily solve Einstein's equation. So it is referred to as the off-shell temperature. In fact only the extremal points on the Gibbs free energy landscape (\(G_L\)) represent the true black hole phases that comply with Einstein's field equations. In this framework, local maxima and minima indicate unstable and locally stable black hole phases, respectively.\\

Another form of landscape energy used to investigate phase transitions based on thermal fluctuations is the thermal potential model introduced in \cite{23},
\begin{equation}\label{(4)}
U =\int \left(T_{H}\ - T \ \right)d S,
\end{equation}
Here, \(T_H\) is the Hawking temperature, \(T\) is the 'off-shell' temperature, and \(S\) is entropy, which acts as a variable. In thermodynamics, the intensity of thermal motion is measured by the product of temperature and entropy. Therefore, the thermal potential serves as an approximate measure of how much all possible states within the canonical ensemble deviate from the true black hole state.
\section{AdS charged black hole with Kaniadakis statistics}
The general metric in 4-dimensional charged AdS black holes is given by\cite{35}:
\begin{equation}\label{(5)}
ds^2 = -f(r)dt^2 + \frac{dr^2}{f(r)} + r^2 d\Omega^2,
\end{equation}
where \( d\Omega^2 = d\theta^2 + \sin^2 \theta d\phi^2 \) is the angular part of the metric on the two-sphere. Here, we have have:
\begin{equation}\label{(6)}
f(r) = 1 - \frac{2M}{r} + \frac{Q^2}{r^2} + \frac{r^2}{l^2},
\end{equation}
where \( M \), \( Q \), and \( l \) are the normalized mass, electric charge, and AdS radius of the black hole, respectively. Also, the cosmological constant is $\Lambda = -\frac{3}{l^2}$.
Clearly, for \( Q = 0 \) and \( l \gg r \), the equation reduces to the well-known Schwarzschild metric. Additionally, the normalized event horizon \( r_+ \) of the geometry corresponds to the largest root of \( f(r) = 0 \). One can use this solution to express the black hole mass as:
\begin{equation}\label{(7)}
M = \frac{r_+}{2} \left( 1 + \frac{Q^2}{r_+^2} + \frac{r_+^2}{l^2} \right).
\end{equation}
The normalized surface area \( A_{\text{BH}} \) of the black hole horizon is $A_{\text{BH}} = 4\pi r_+^2.$
Returning to our system of Planck units, the Bekenstein-Hawking entropy based on classical Boltzmann-Gibbs statistics is:
\begin{equation}\label{(8)}
S = \frac{A_{\text{BH}}}{4}.
\end{equation}
Notice that, according to the set of units we are using, this is a dimensionless measure of the black hole horizon entropy, rescaled by the number of bits associated with a Planck-size area. As discussed in the Introduction, due to the area scaling of black hole entropy, arguments from multiple perspectives suggest that Boltzmann-Gibbs statistics may not be the appropriate context for studying the thermodynamics of black holes. In particular, in a relativistic scenario, the entropy is expected to be generalized to Kaniadakis entropy, which we rewrite here by dropping the index \( \kappa \):
\begin{equation}\label{(9)}
S_{\kappa} = \frac{1}{\kappa}\sinh(\kappa S_{BH}).
\end{equation}
Some comments are in order: first, it should be stressed that, although the Bekenstein-Hawking entropy is commonly used in relativistic theory, it is essentially classical. According to the maximum entropy principle, it is maximized when a given system in thermodynamic equilibrium is described by the Maxwell-Boltzmann distribution, which has classical Boltzmann-Gibbs statistics as its natural framework. To learn more about the mentioned model, you can see \cite{35}. Here, we will apply the concepts discussed in the previous section to the current model and then compare its results with other works in the literature. For the primary quantities of this model, namely the mass (M) and the Hawking temperature (T), we obtain the following\cite{35}:
\begin{equation}\label{(10)}
M=\frac{3 \pi  q^{2} \kappa^{2}+8 \mathrm{arcsinh}\! \left(\kappa  S \right)^{2} P +3 \kappa  \mathrm{arcsinh}\! \left(\kappa  S \right)}{6 \sqrt{\pi}\, \kappa^{\frac{3}{2}} \sqrt{\mathrm{arcsinh}\! \left(\kappa  S \right)}},
\end{equation}
and,
\begin{equation}\label{(12)}
T =\frac{-\pi  q^{2} \kappa^{2}+8 \mathrm{arcsinh}\! \left(\kappa  S \right)^{2} P +\kappa  \mathrm{arcsinh}\! \left(\kappa  S \right)}{4 \sqrt{\kappa}\, \sqrt{\kappa^{2} S^{2}+1}\, \sqrt{\pi}\, \mathrm{arcsinh}\! \left(\kappa  S \right)^{\frac{3}{2}}},
\end{equation}
Now, based on equations (3) and (4) for the Gibbs and thermal potentials of this model, we obtain the following,
\begin{equation}\label{(11)}
\begin{split}
&U =\frac{3 \pi  q^{2} \kappa^{2}+8 {\mathrm{arcsinh}\! \left(\kappa  \left(\pi  r^{2}+\frac{1}{6} \pi^{3} \kappa^{2} r^{6}\right)\right)}^{2} P +3 \kappa  \mathrm{arcsinh}\! \left(\kappa  \left(\pi  r^{2}+\frac{1}{6} \pi^{3} \kappa^{2} r^{6}\right)\right)}{6 \sqrt{\pi}\, \kappa^{\frac{3}{2}} \sqrt{\mathrm{arcsinh}\! \left(\kappa  \left(\pi  r^{2}+\frac{1}{6} \pi^{3} \kappa^{2} r^{6}\right)\right)}}\\&-T \left(\pi  r^{2}+\frac{1}{6} \pi^{3} \kappa^{2} r^{6}\right),
\end{split}
\end{equation}
and
\begin{equation}\label{(12)}
\begin{split}
&\widetilde{G }=4\bigg[\mathrm{arcsinh}\times \big(\frac{\kappa  \pi  r^{2} (\pi^{2} \kappa^{2} r^{4}+6)}{6}\big) \bigg(\frac{3 \pi  q^{2} \kappa^{2}}{8}+\mathrm{arcsinh}\times (\frac{\kappa  \pi  r^{2} (\pi^{2} \kappa^{2} r^{4}+6)}{6})^{2} P \\&+\frac{3 \kappa  \mathrm{arcsinh}\big(\frac{\kappa  \pi  r^{2} (\pi^{2} \kappa^{2} r^{4}+6)}{6}\big)}{8}\bigg)\times \sqrt{\kappa^{6} \pi^{6} r^{12}+12 \kappa^{4} \pi^{4} r^{8}+36 \pi^{2} \kappa^{2} r^{4}+36}\\&-\frac{3 \pi  (\pi^{2} \kappa^{2} r^{4}+6) \kappa  \bigg[-\frac{\pi  q^{2} \kappa^{2}}{8}+\mathrm{arcsinh}\big(\frac{\kappa  \pi  r^{2} (\pi^{2} \kappa^{2} r^{4}+6)}{6}\big)^{2} P +\frac{\kappa  \mathrm{arcsinh}\big(\frac{\kappa  \pi  r^{2} (\pi^{2} \kappa^{2} r^{4}+6)}{6}\big)}{8}\bigg] r^{2}}{2}\bigg]\\&\bigg/3 \kappa^{\frac{3}{2}} \mathrm{arcsinh}\! \big(\frac{\kappa  \pi  r^{2} (\pi^{2} \kappa^{2} r^{4}+6)}{6}\big)^{\frac{3}{2}} \sqrt{\pi}\times \sqrt{\kappa^{6} \pi^{6} r^{12}+12 \kappa^{4} \pi^{4} r^{8}+36 \pi^{2} \kappa^{2} r^{4}+36}.
\end{split}
\end{equation}
Next, we will explore a special state formed based on arbitrary parameter settings. By assigning specific values to the parameters and performing some calculations for the critical quantities, we obtain the following results,
\begin{equation}\label{(13)}
T_{c}=\frac{\sqrt{6}}{18 \pi  q}+3 \sqrt{6}\, \pi  q^{3} \kappa^{2},
\end{equation}
\begin{equation}\label{(14)}
P_{c}=\frac{1}{96 \pi  q^{2}}+2 \pi  q^{2} \kappa^{2},
\end{equation}
and
\begin{equation}\label{(15)}
r_{c}=\frac{3^{\frac{1}{3}} 4^{\frac{2}{3}} \left(\frac{8 \sqrt{6}\, \pi  q^{3}+1728 \sqrt{6}\, \pi^{3} q^{7} \kappa^{2}}{\pi}\right)^{\frac{1}{3}}}{4},
\end{equation}
So with respect to $0.25, \kappa= 0.01$, we can obtain,
$$r_c = 0.6125423725,\hspace{0.25cm}
P_c = 0.05309091760,\hspace{0.25cm}
T_c = 0.1733020275,\hspace{0.25cm}
U_c = 0.2041147009,\hspace{0.25cm}
G_c = 0.2041147037$$
Following the standard procedure, we convert the parameters into dimensionless form. This not only facilitates comparison with other studies but also adds convenience,
\begin{equation}\label{(16)}
t = \frac{T}{T_{c}},\hspace{0.7cm}p = \frac{P}{P_{c}},\hspace{0.7cm}x = \frac{r}{r_{c}},\hspace{0.7cm}\widetilde{G} = \frac{G}{G_{c}},\hspace{0.7cm}u = \frac{U}{U_{c}}.
\end{equation}

\begin{figure}[h!]
 \begin{center}
 \subfigure[]{
 \includegraphics[height=6cm,width=8cm]{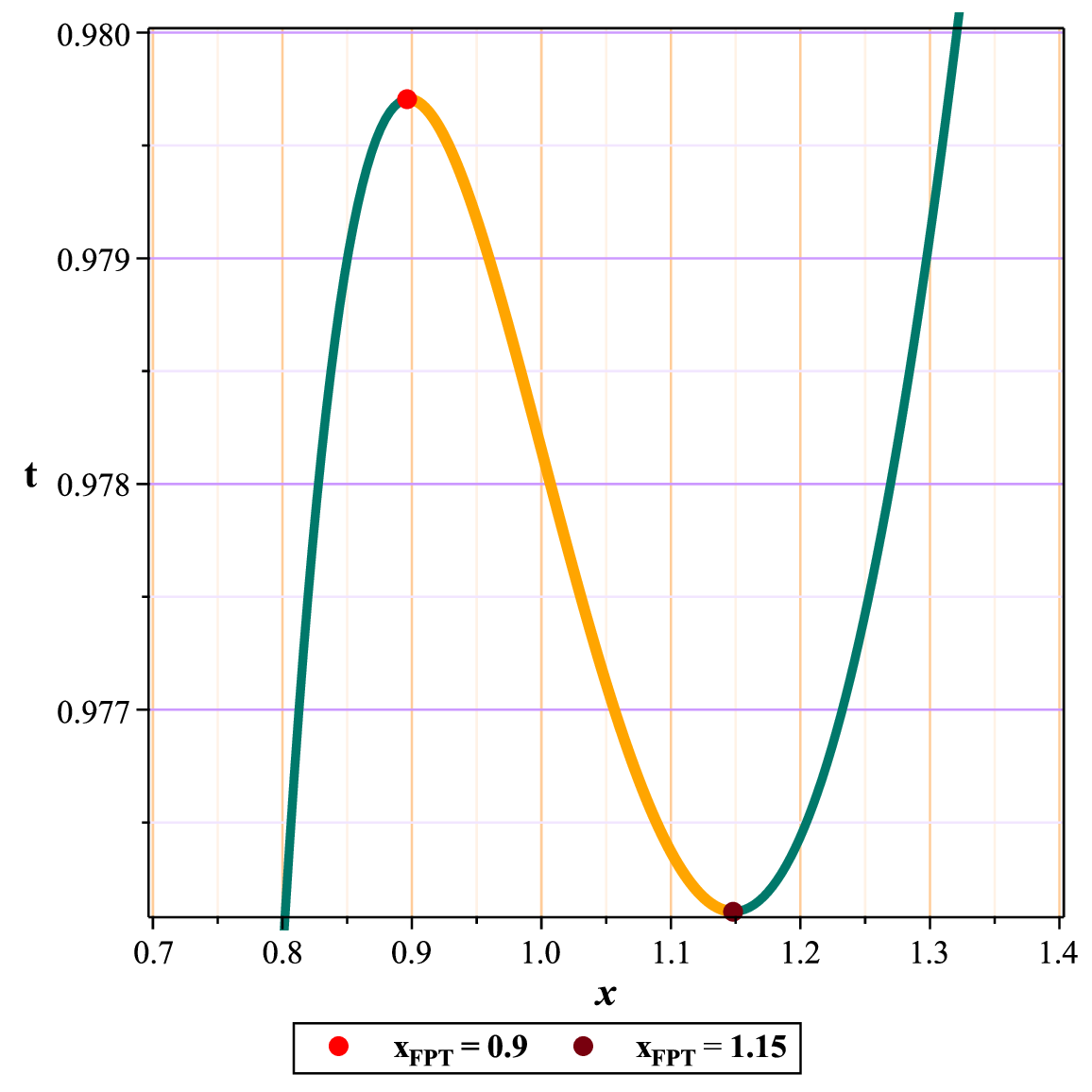}
 \label{fig1a}}
 \subfigure[]{
 \includegraphics[height=6cm,width=8cm]{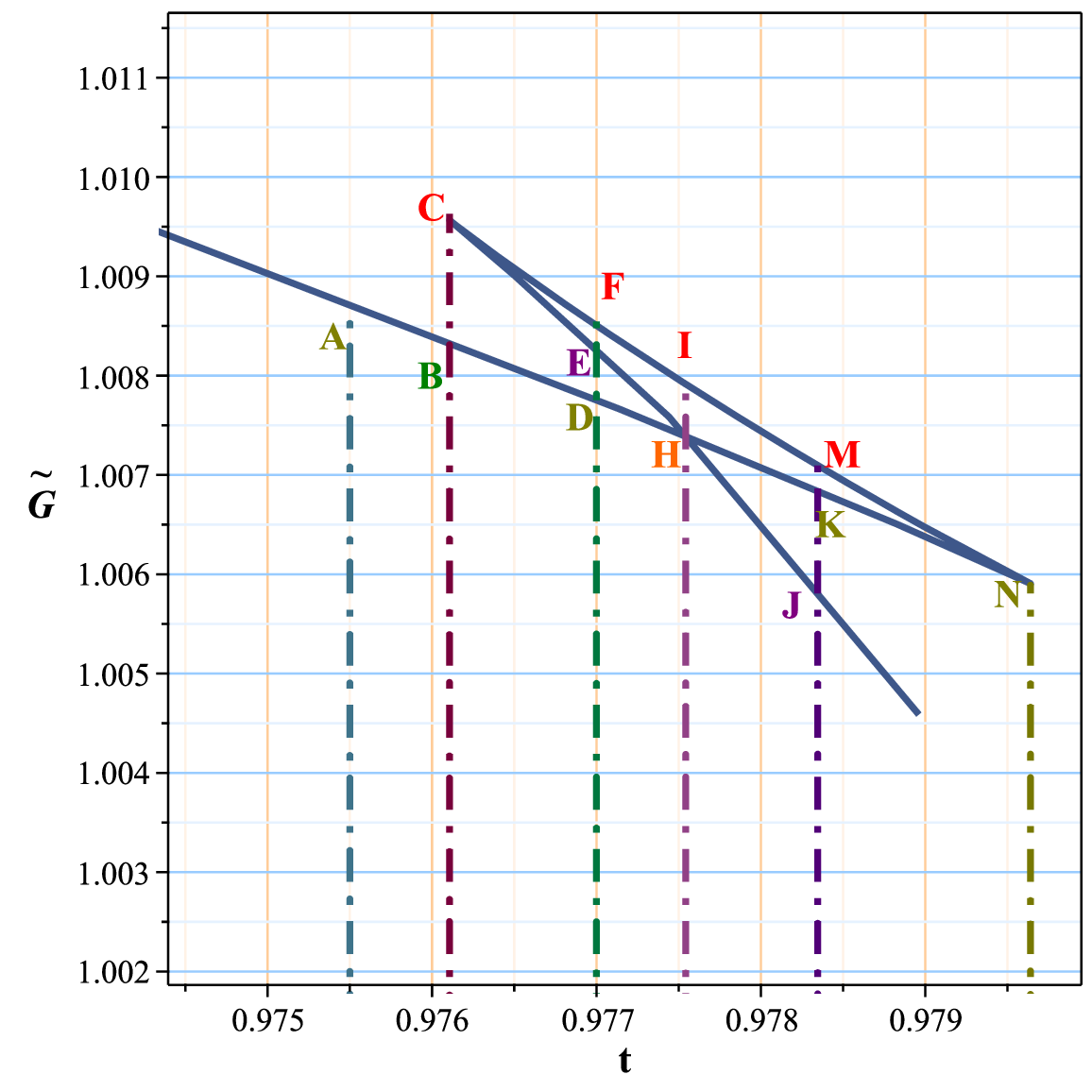}
 \label{fig1b}}
 \caption{\small{The Figure of temperature ($T$) against the free parameters ($x$) is shown in Figure 1(a), while Figure 1(b) depicts the Gibbs free energy as a function of temperature.}}
 \label{fig1}
 \end{center}
 \end{figure}

\begin{figure}[h!]
 \begin{center}
 \subfigure[]{
 \includegraphics[height=5cm,width=6cm]{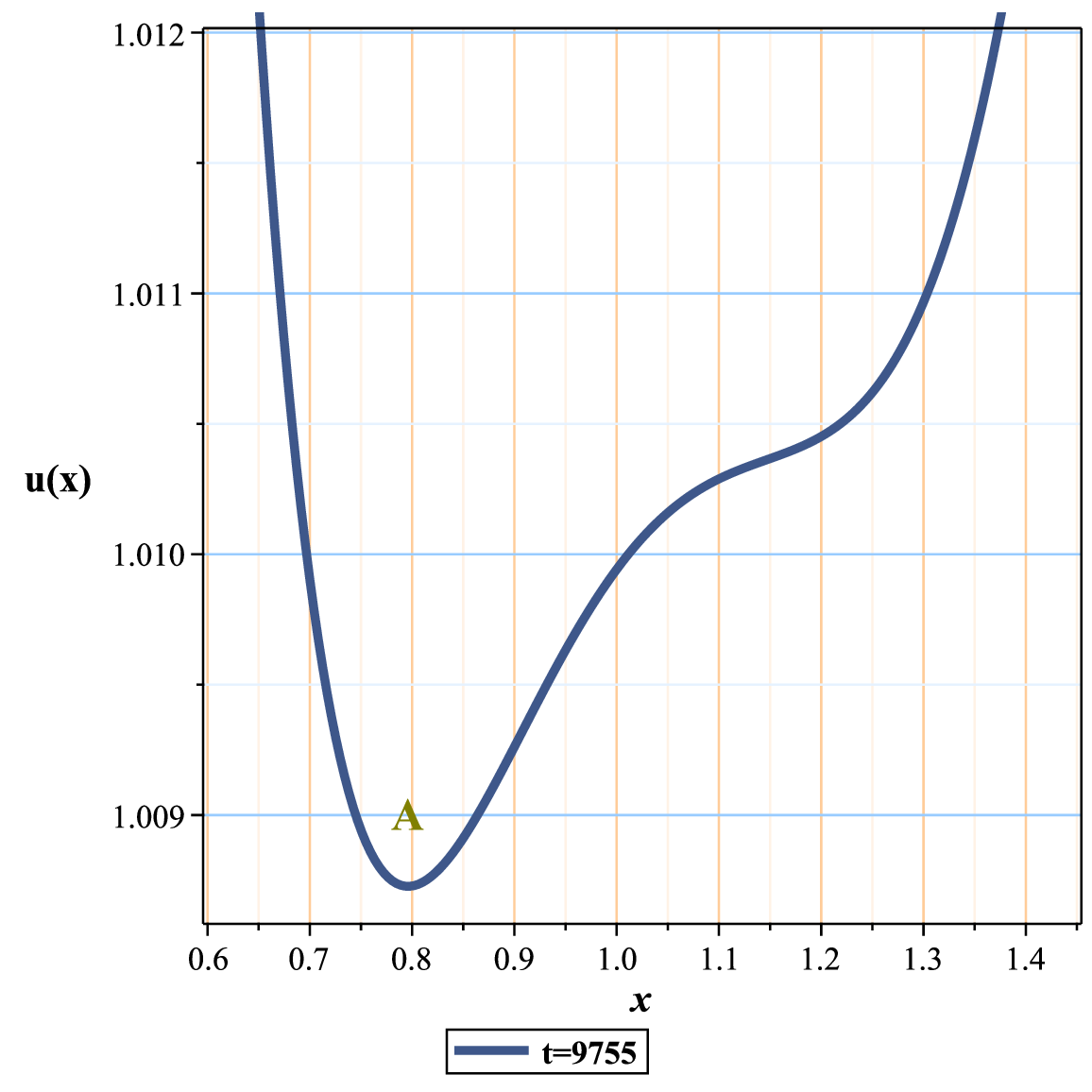}
 \label{fig2a}}
 \subfigure[]{
 \includegraphics[height=5cm,width=6cm]{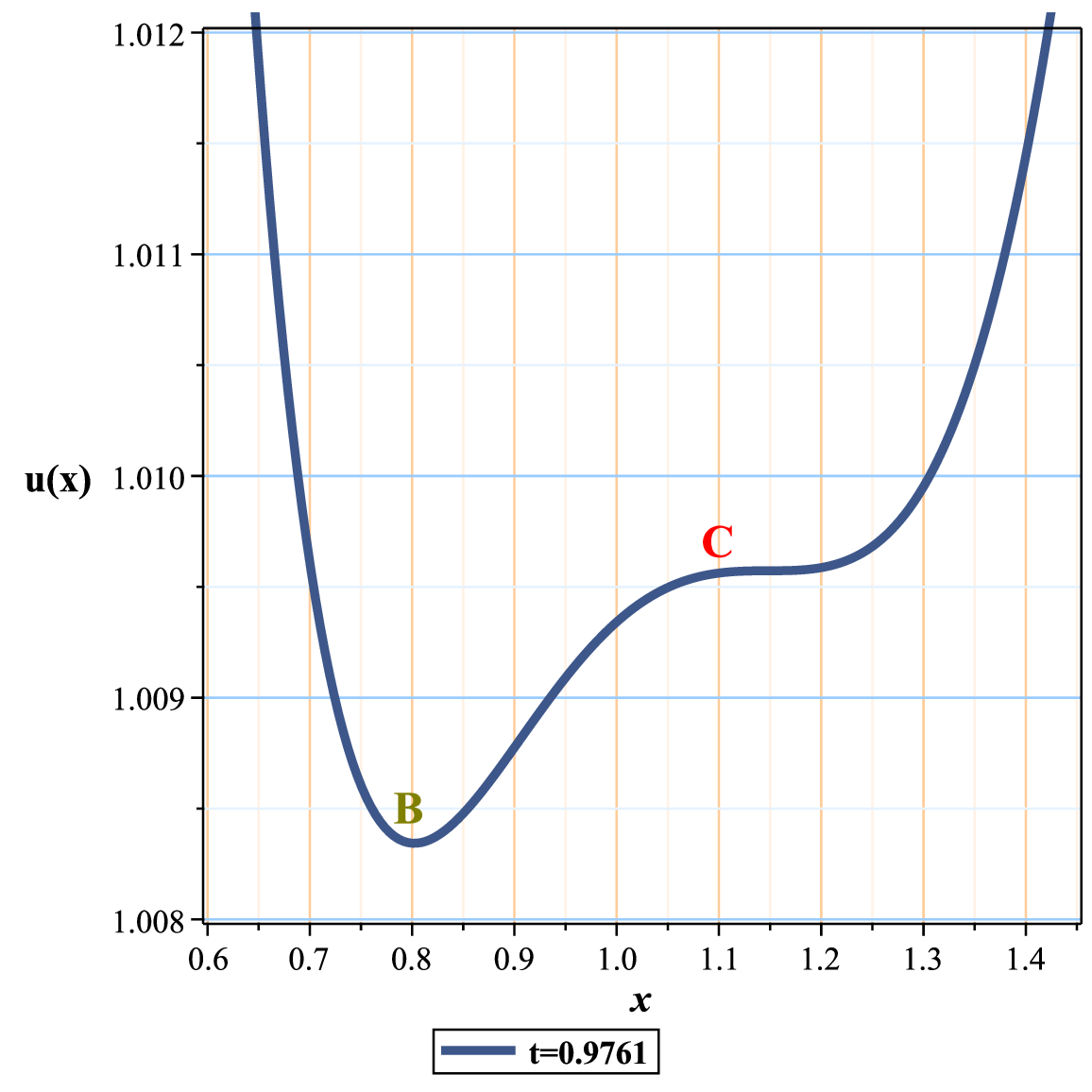}
 \label{fig2b}}
 \subfigure[]{
 \includegraphics[height=5cm,width=6cm]{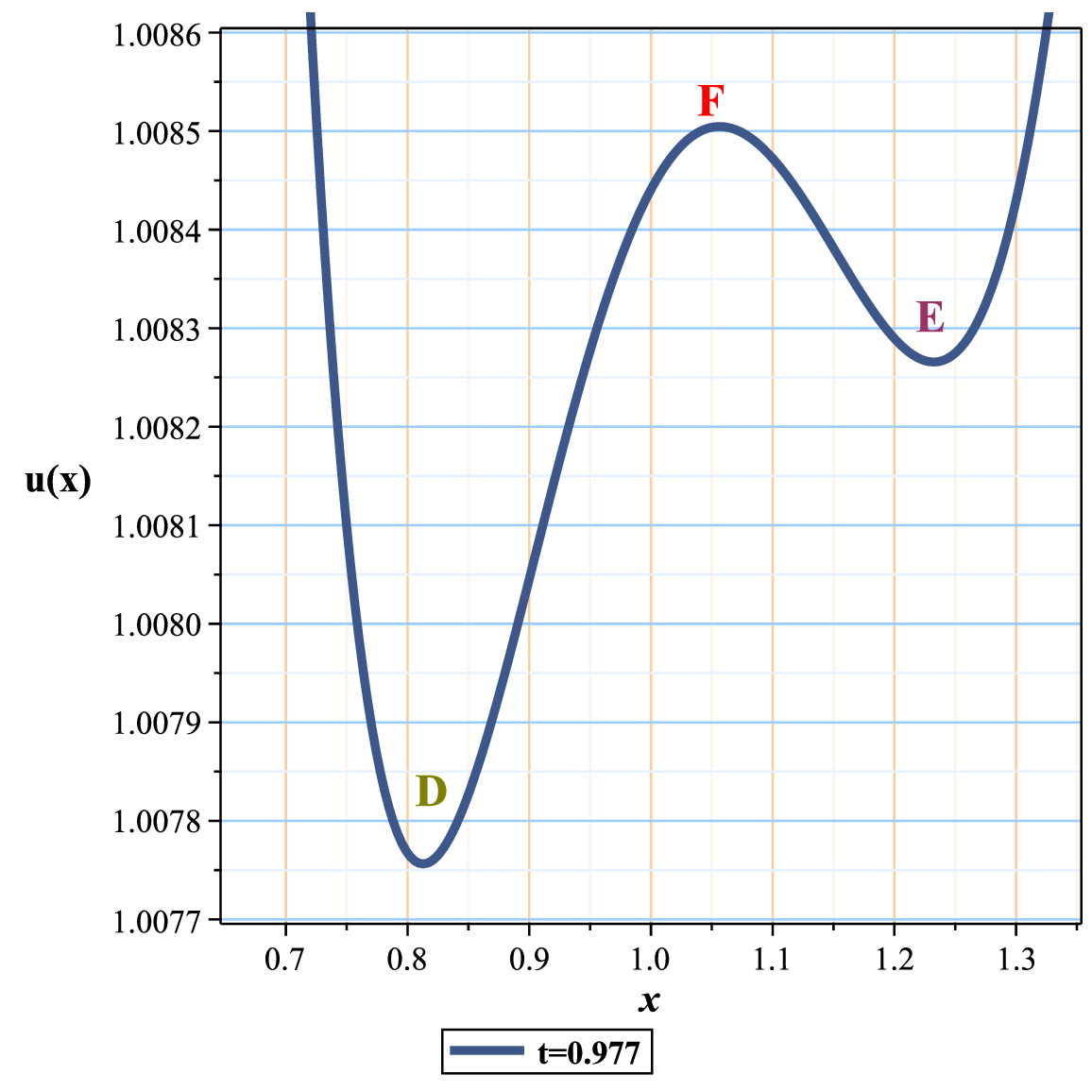}
 \label{fig2c}}
 \subfigure[]{
 \includegraphics[height=5cm,width=6cm]{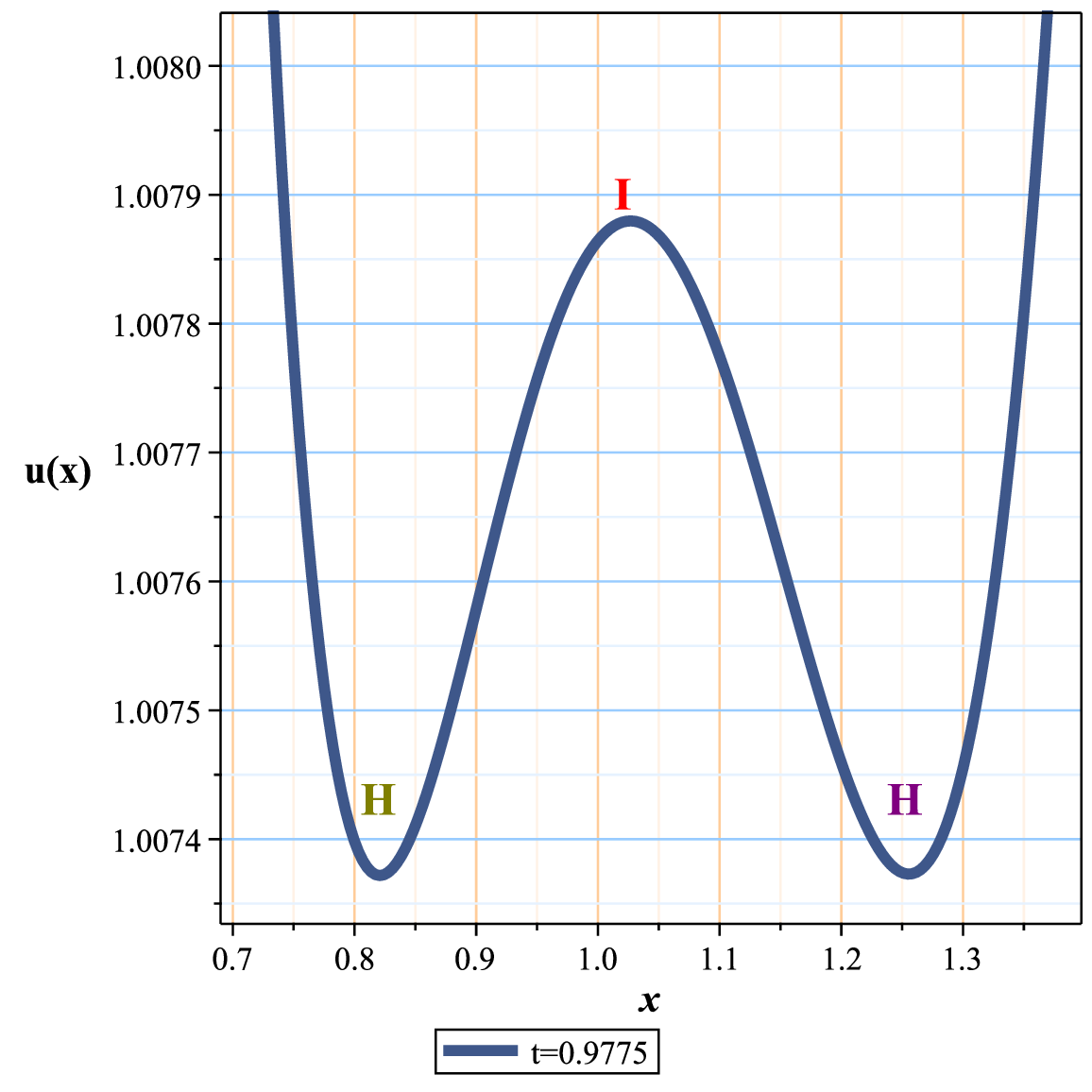}
 \label{fig2d}}
 \subfigure[]{
 \includegraphics[height=5cm,width=6cm]{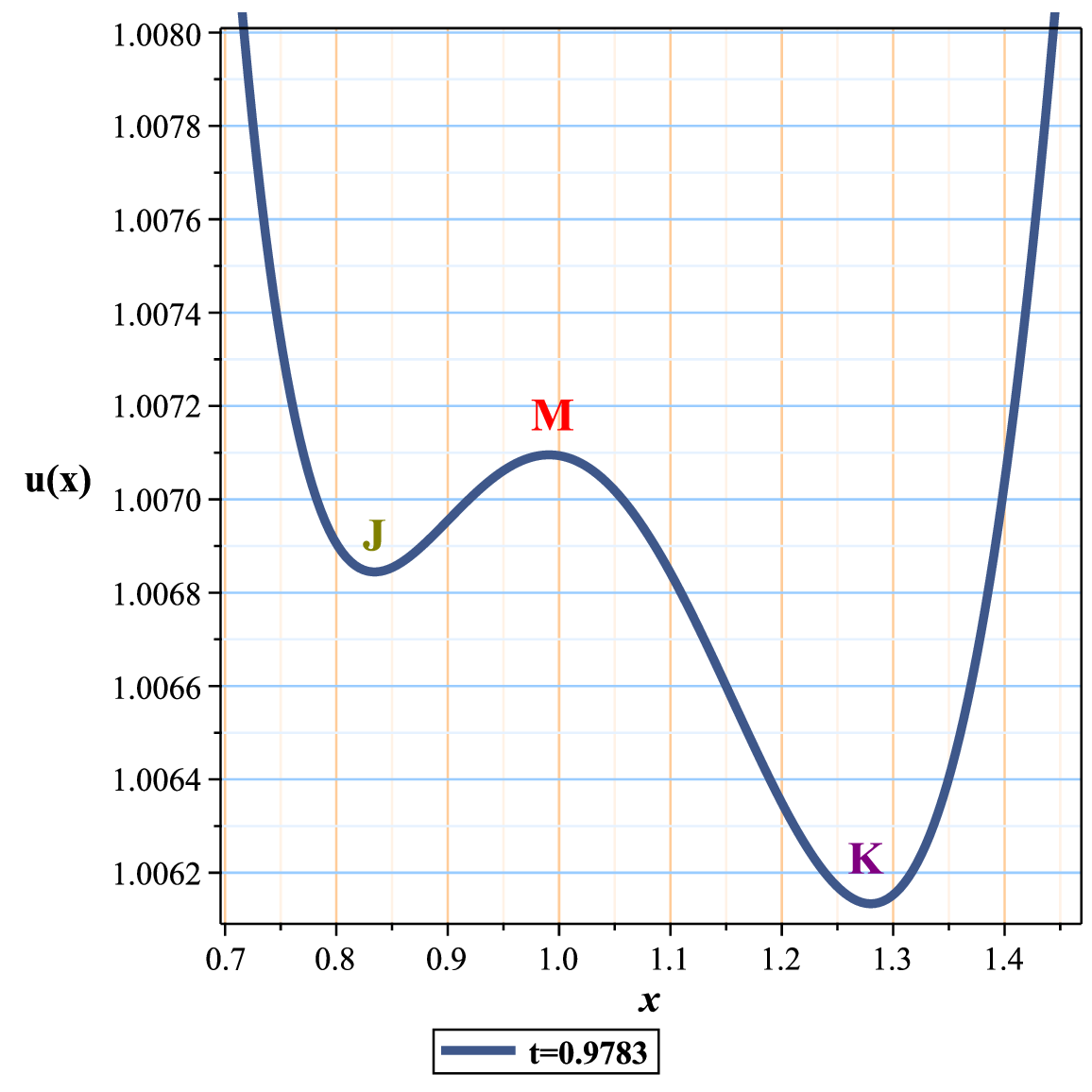}
 \label{fig2e}}
 \subfigure[]{
 \includegraphics[height=5cm,width=6cm]{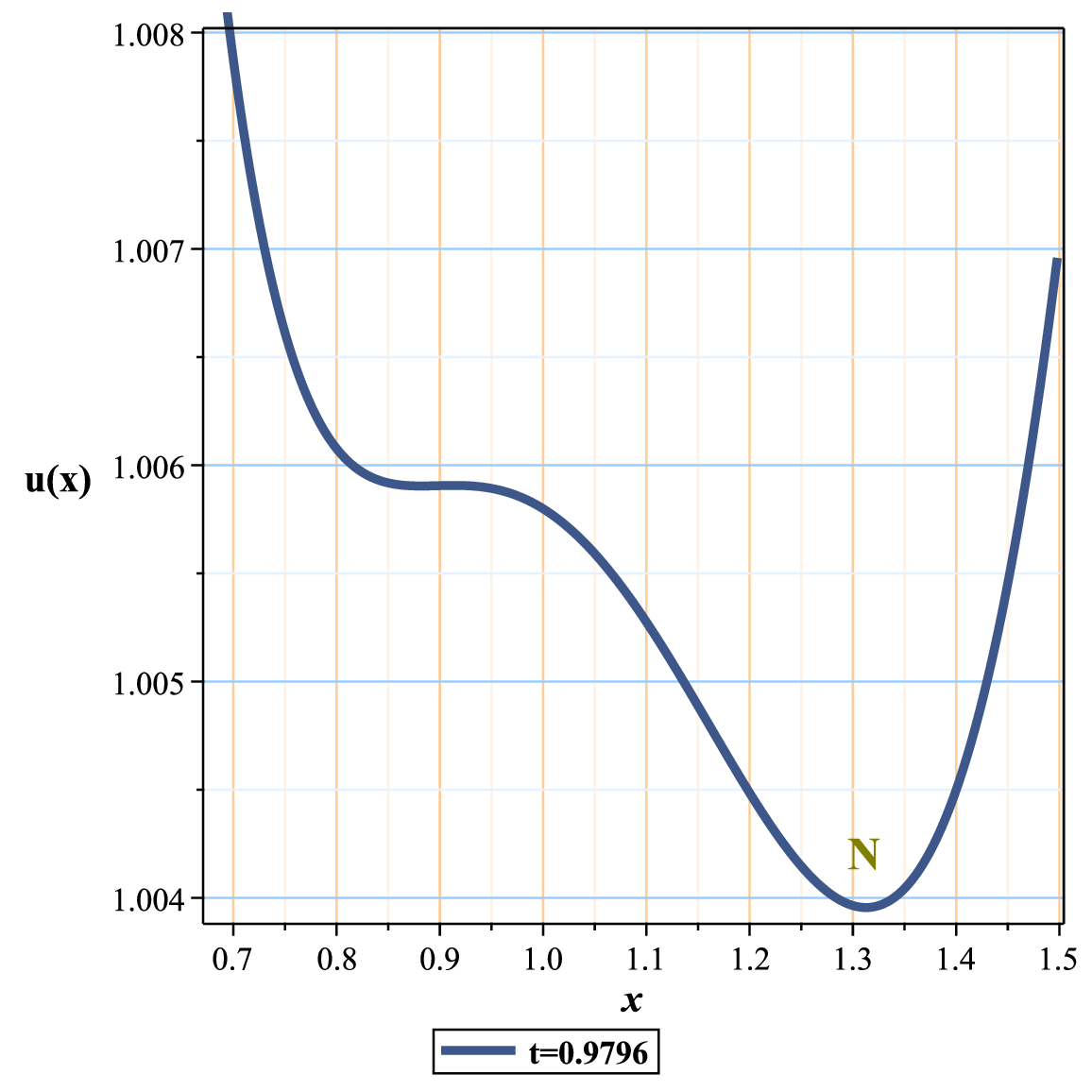}
 \label{fig2f}}
 \caption{\small{Sequence of energy behavior as a function of x at varying temperatures}}
 \label{fig1}
 \end{center}
 \end{figure}

\begin{figure}[h!]
 \begin{center}
 \subfigure[]{
 \includegraphics[height=5cm,width=6cm]{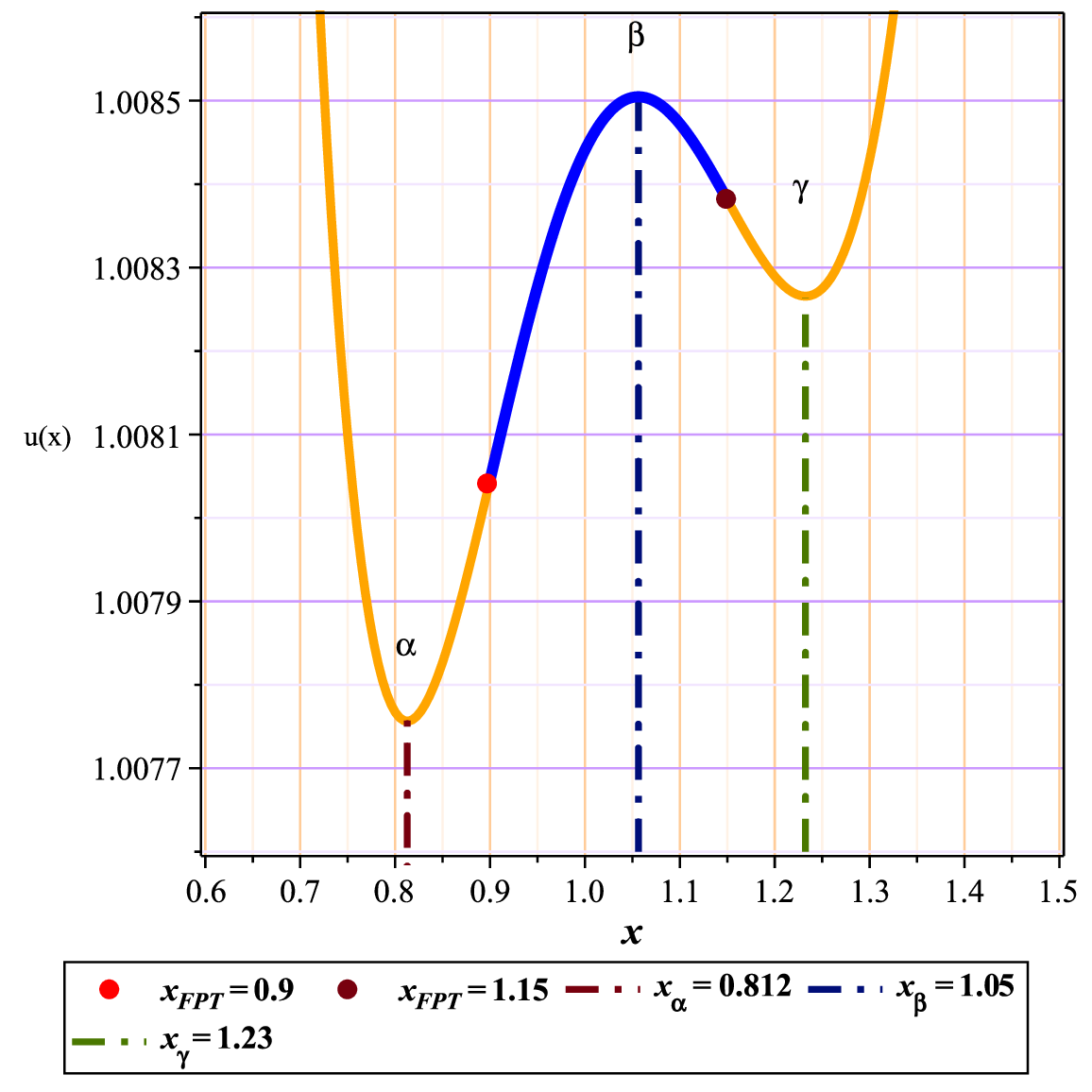}
 \label{fig3a}}
 \subfigure[]{
 \includegraphics[height=5cm,width=6cm]{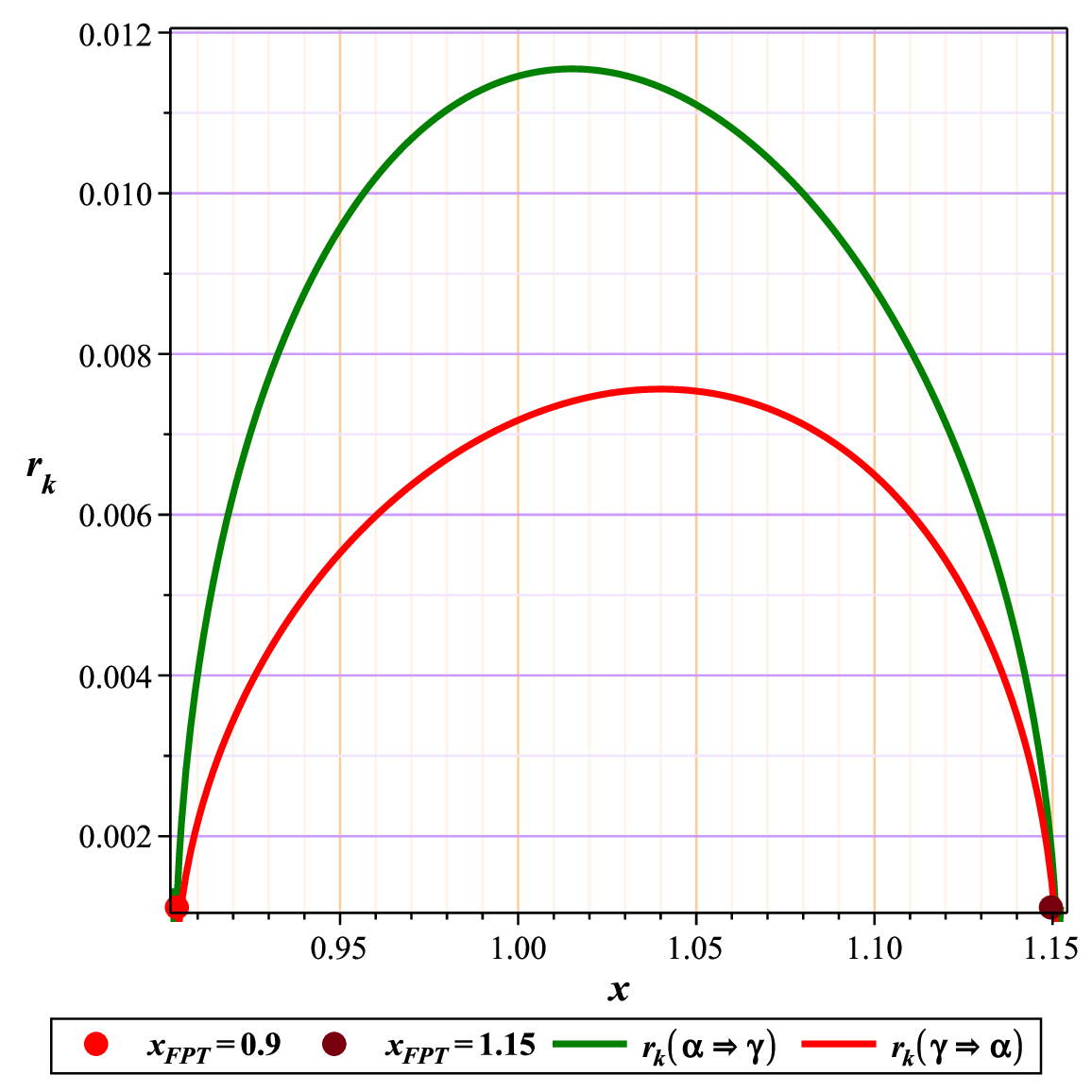}
 \label{fig3b}}
 \subfigure[]{
 \includegraphics[height=5cm,width=6cm]{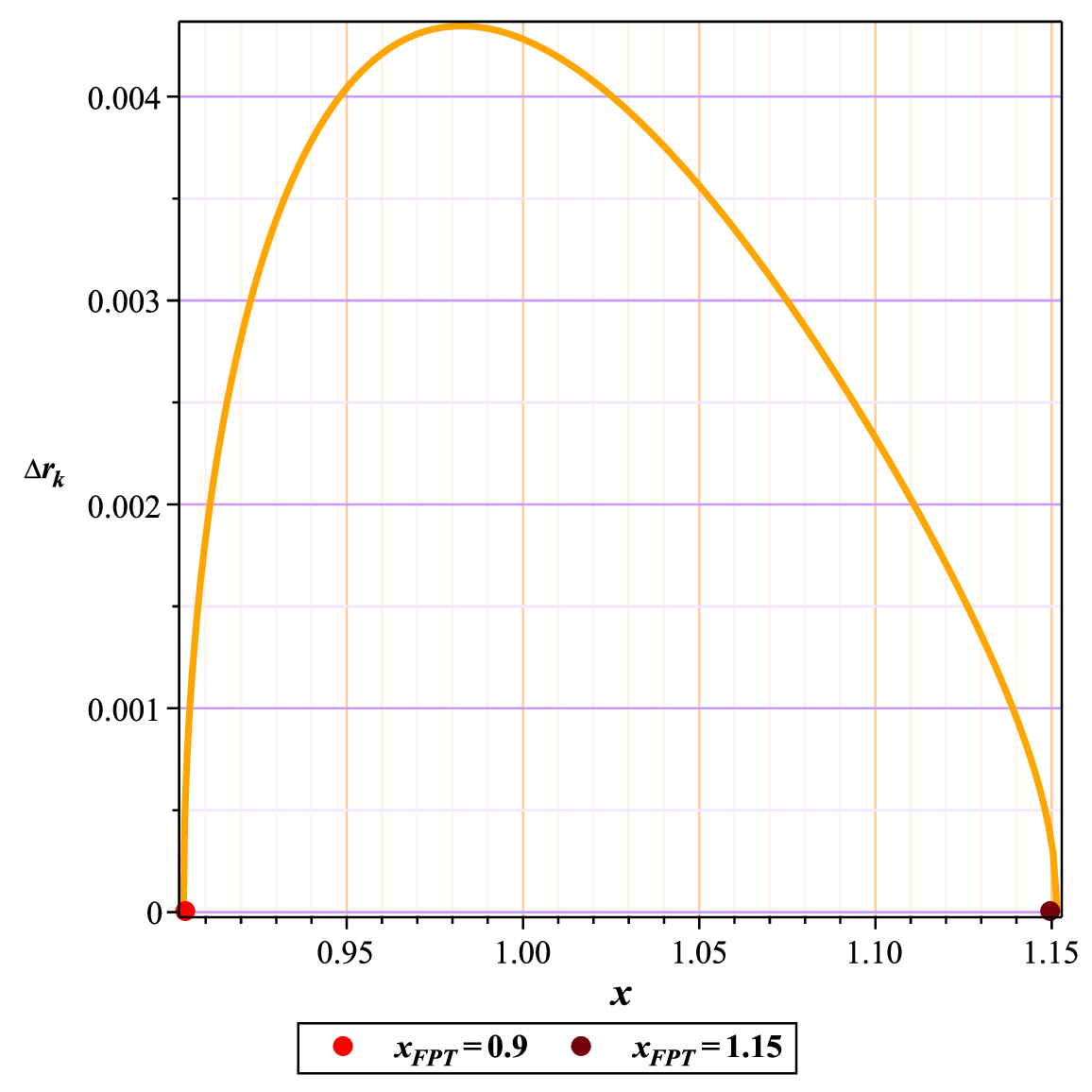}
 \label{fig3c}}
 \caption{\small{Figure 3a illustrates the first-order phase transition region with respect to t and x. It presents the plot of u(r) versus x, considering free parameters and the FPT points. Figure 3b compares $r_k$ as a function of x for two different conditions ($\alpha\rightarrow\gamma$) and ($\gamma\rightarrow\alpha$). It also includes the coordinates of the contact point. Figure 3c depicts $r_k$ (the difference between $\Delta r_k$ $r_k$($\alpha\rightarrow\gamma$)-$r_k$($\gamma\rightarrow\alpha$)) as a function of x}}
 \label{fig3}
 \end{center}
 \end{figure}

\begin{figure}[h!]
 \begin{center}
 \subfigure[]{
 \includegraphics[height=5cm,width=6cm]{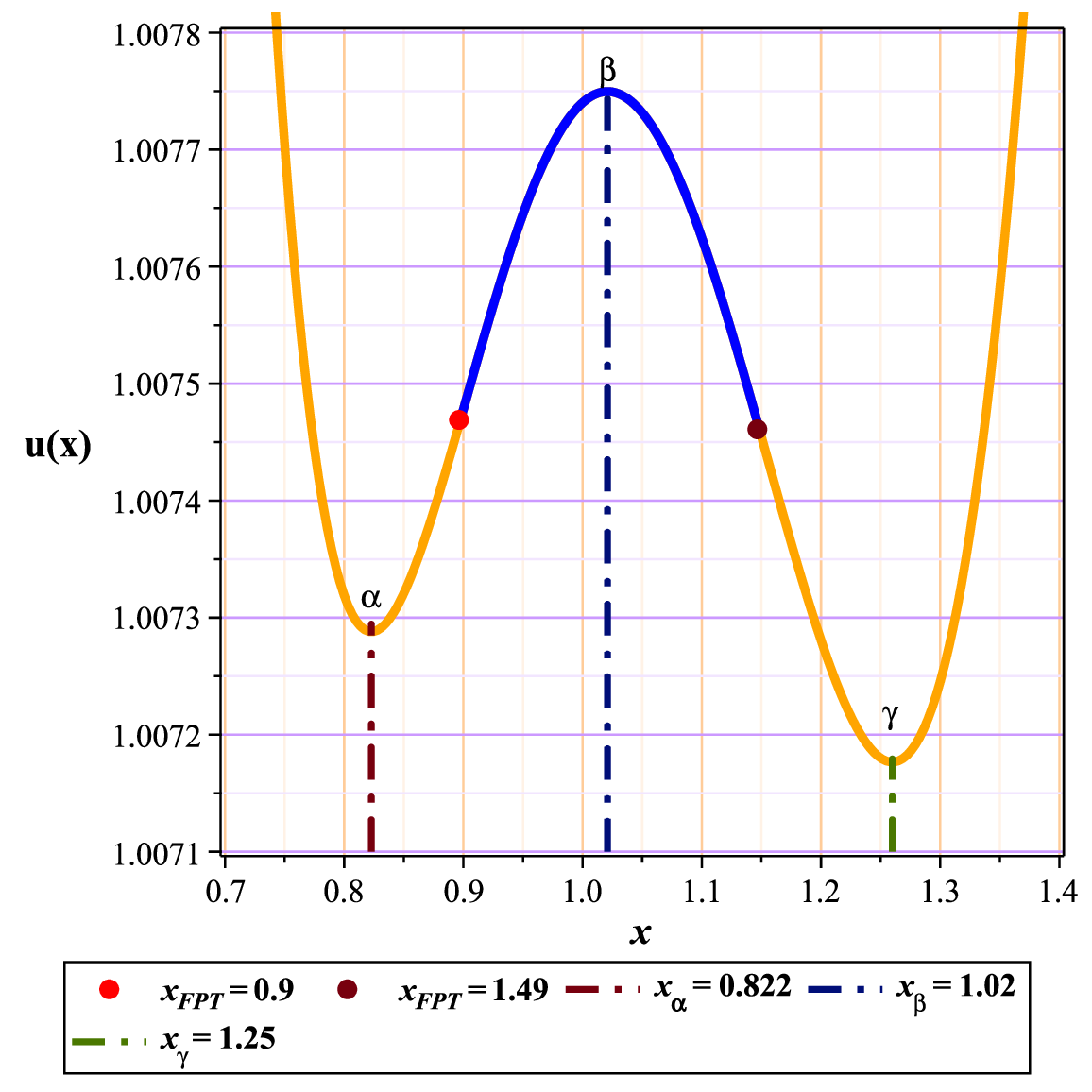}
 \label{fig4a}}
 \subfigure[]{
 \includegraphics[height=5cm,width=6cm]{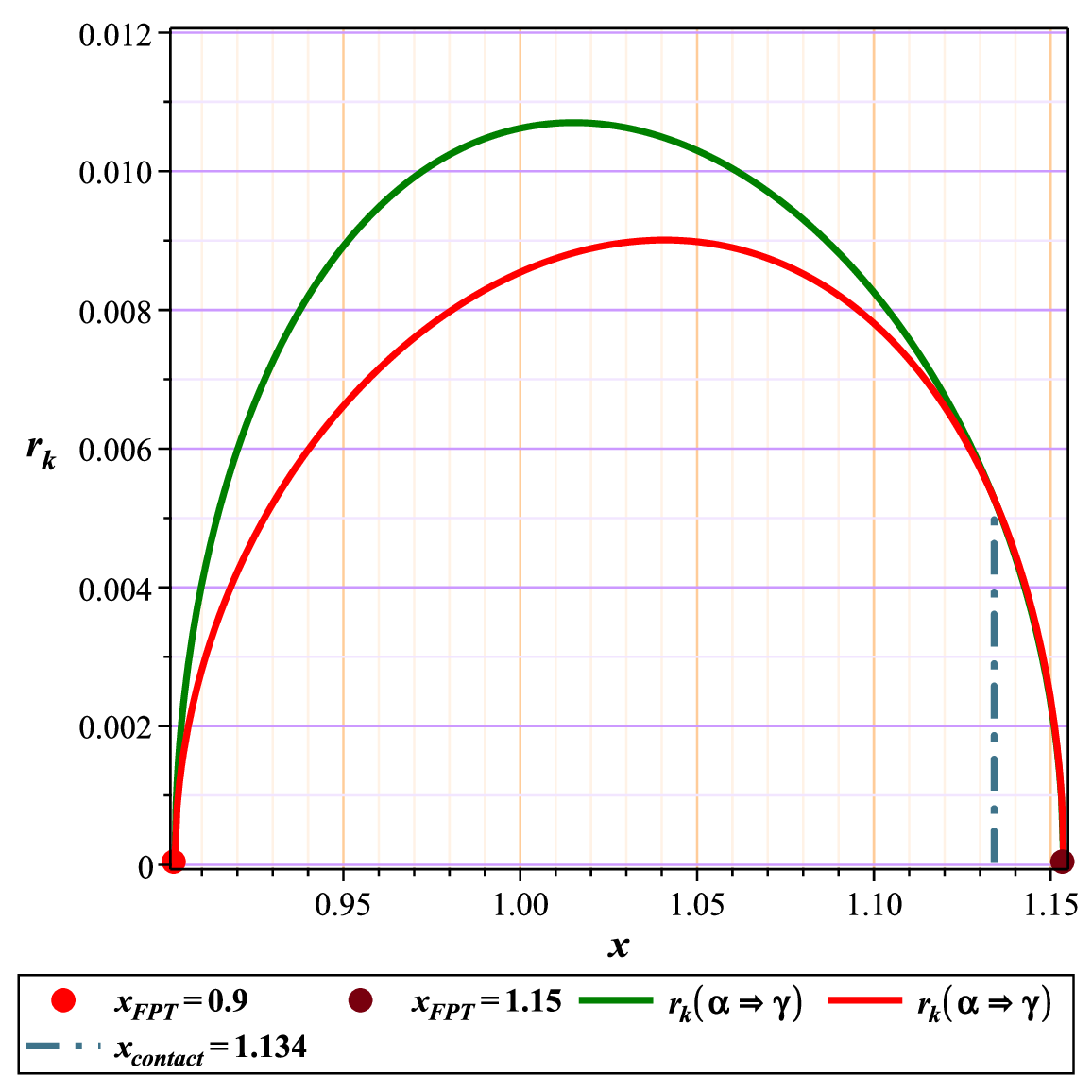}
 \label{fig4b}}
 \subfigure[]{
 \includegraphics[height=5cm,width=6cm]{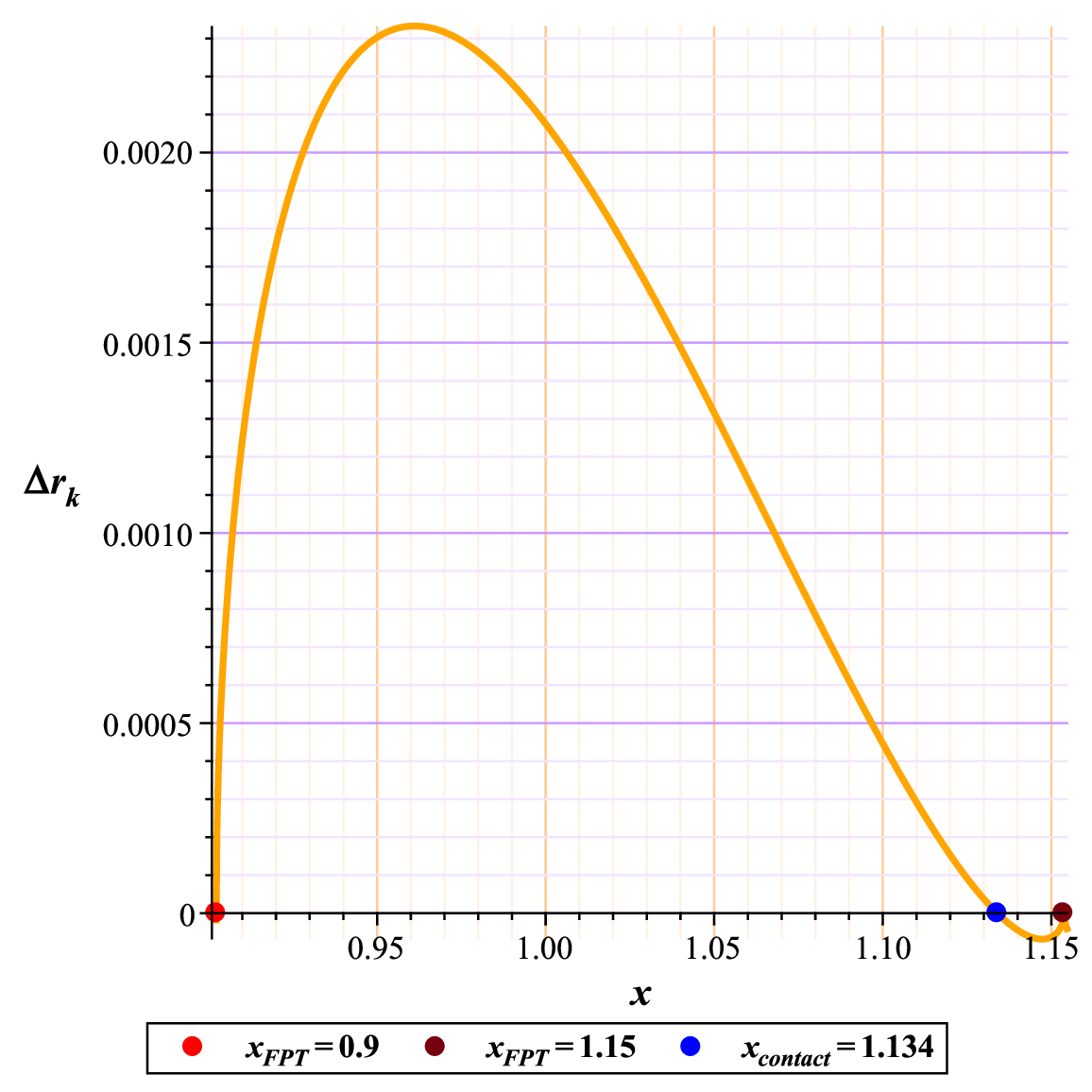}
 \label{fig4c}}
 \caption{\small{Figure 4 consists of the following parts:
- (4a): It shows the plot of u(r) versus x, considering free parameters and the FPT points.
- (4b): This part compares $r_k$ as a function of x for two distinct conditions ($\alpha\rightarrow\gamma$) and ($\gamma\rightarrow\alpha$). It also includes the coordinates of the contact point.
- (4c): This section presents $r_k$ as a function of x}}
 \label{fig4}
 \end{center}
 \end{figure}

 \begin{figure}[h!]
 \begin{center}
 \subfigure[]{
 \includegraphics[height=5cm,width=6cm]{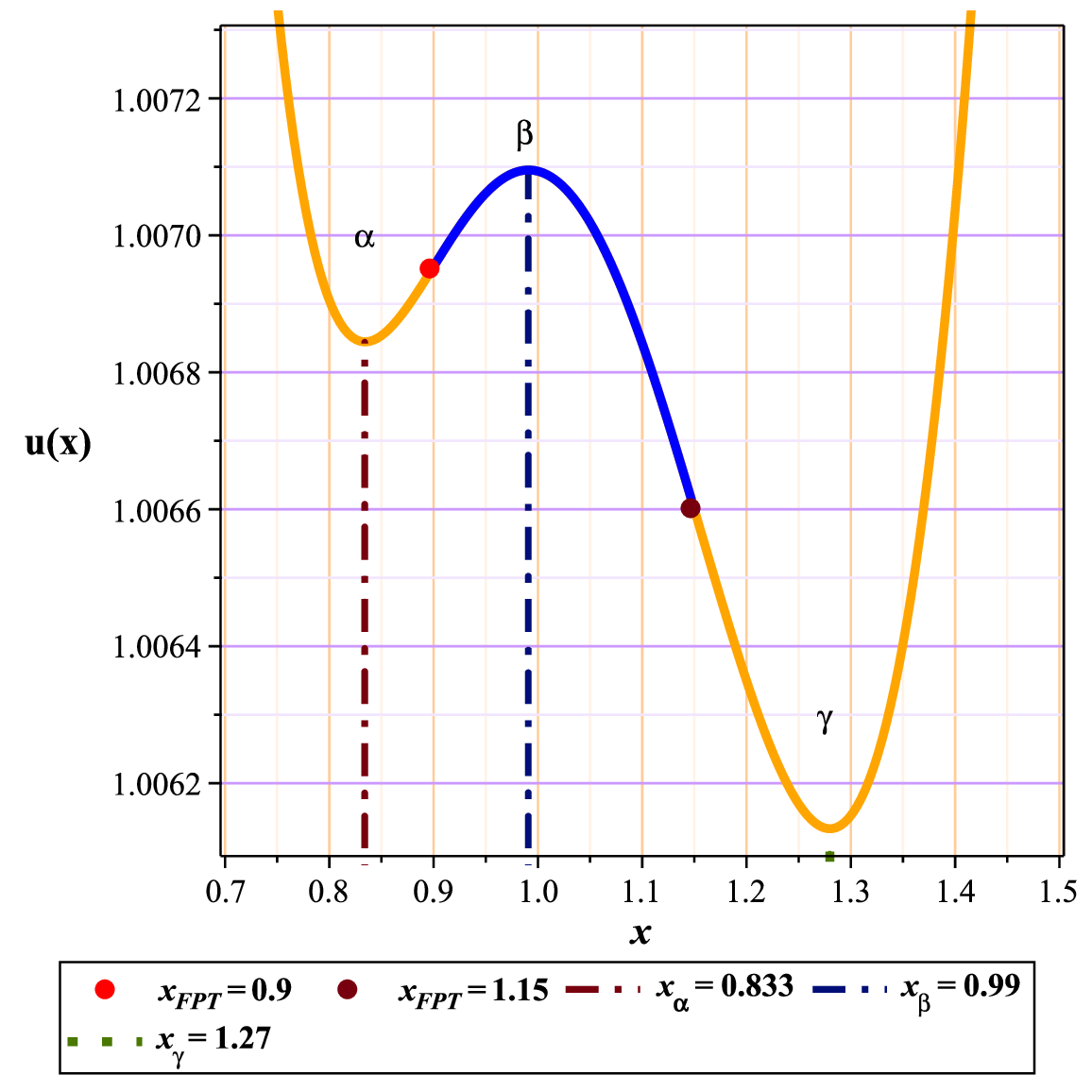}
 \label{fig5a}}
 \subfigure[]{
 \includegraphics[height=5cm,width=6cm]{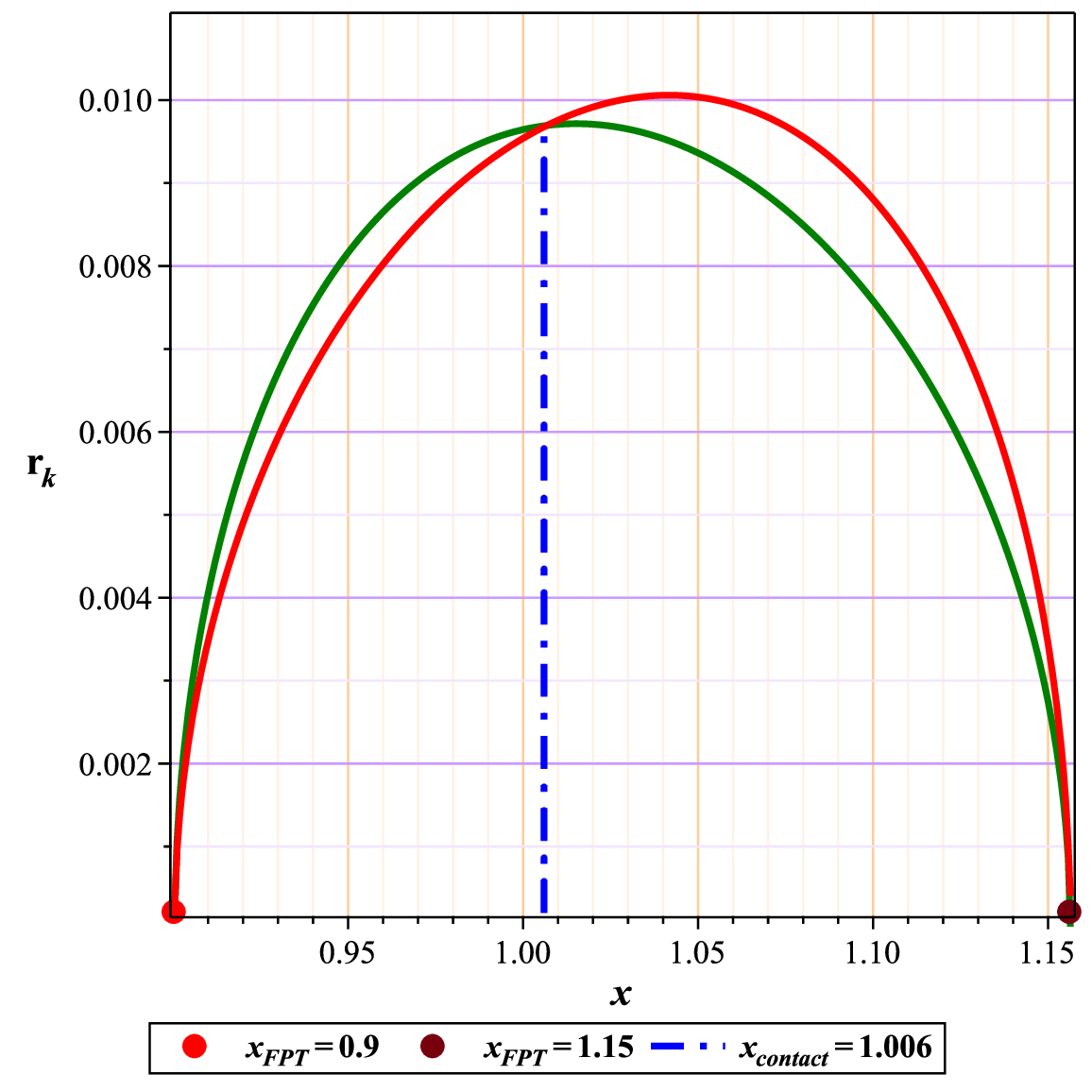}
 \label{fig5b}}
 \subfigure[]{
 \includegraphics[height=5cm,width=6cm]{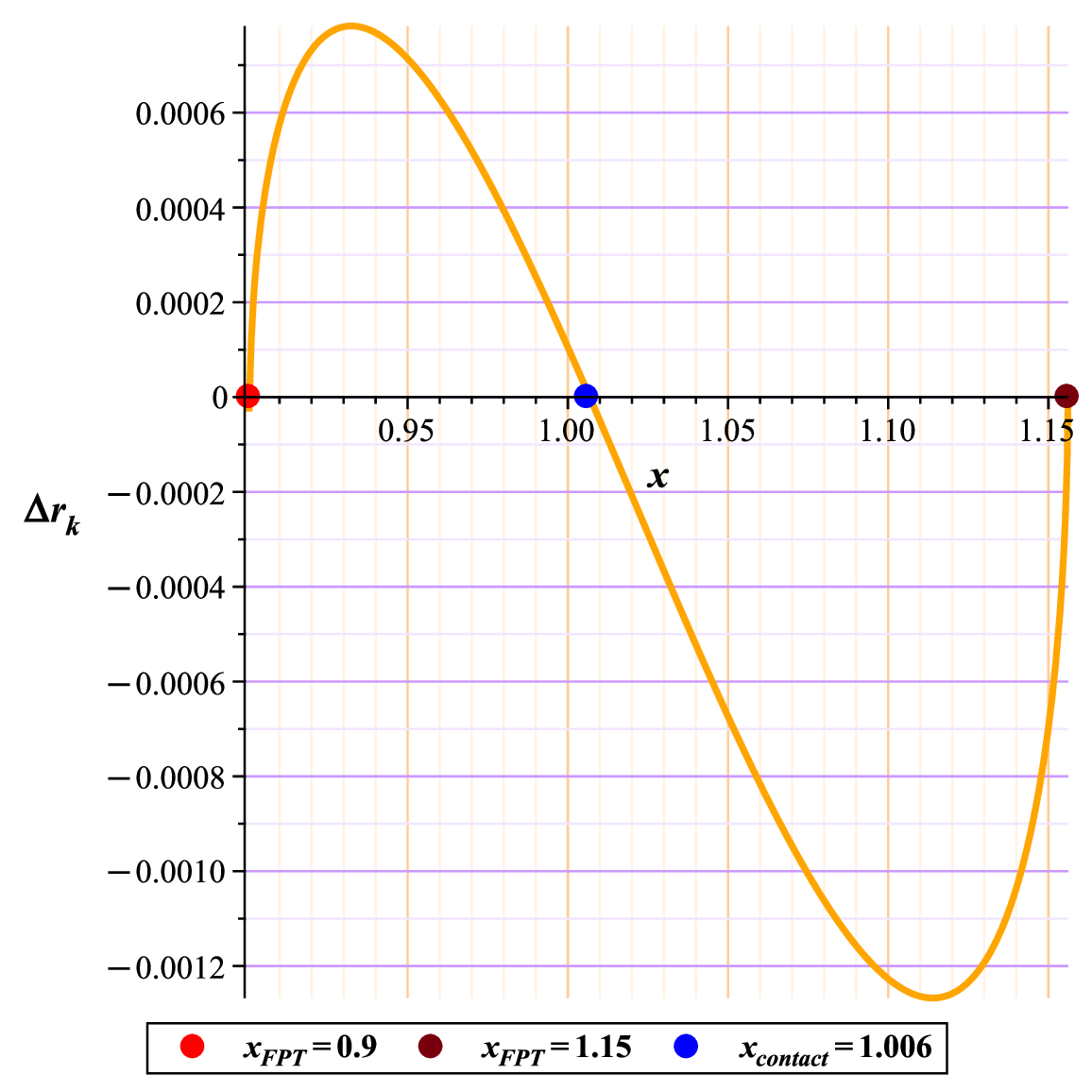}
 \label{fig5c}}
 \caption{\small{(5a): The function u(r) plotted against x for free parameters and the FPT (First Passage Time) points.
(5b): A comparison of $r_k$ with respect to x for transitions from $\alpha$ to $\gamma$ and from $\gamma$ to $\alpha$.
(5c): The change in $r_k$ ($\Delta r_k$) with respect to x.}}
 \label{fig5}
 \end{center}
 \end{figure}

To interpret the behavior of the energy function, we rely on the graph of temperature changes (Fig 1a) and the Gibbs free energy as a function of temperature (Fig 1b). Here are the key insights:
In Fig 1b, we observe three lines: (A-N), (C-J), and (N-C).
The line (A-N) represents the small black hole, while (C-J) represents the large black hole. Both are locally stable with positive heat capacity.
The line (N-C) corresponds to the intermediate black hole, which is unstable due to its negative heat capacity.
Classical principles dictate that systems tend to minimize their energy. In the free energy landscape, minima correspond to stable states, while maxima represent unstable states.
Systems with multiple local minima may appear stable within their immediate vicinity but can be globally unstable relative to the primary minimum.
Even if the system initially occupies these local minima, it eventually progresses toward the ultimate global minimum.
Fig 2a shows a global minimum at point A, belonging to the branch of the small black hole.
As temperature increases (Fig 2b), local extrema B (minimum) and C (maximum) begin to form. However, they remain globally unstable.
Fig 2c depicts fully formed local extrema. Gradually, the system transitions toward a large and stable black hole (Fig 2d).
The isotherm line (H-I) in Fig 1b seemingly has a maximum (I) on the intermediate black hole branch and a global minimum (H).
However, in reality, point H exists on both the small and large black holes, creating a coexistence phase.
Temperature increase strengthens the large black hole and weakens the small one (Fig 2e and 2f).
Second-Order Phase Transition:
When studying the system above the critical temperature (Fig 2g), it follows a path similar to (I-H-J) in Fig 1b.
In this article, we assume that the first-order phase transition occurs from a small black hole to a large black hole.
In Fig 3, we find ourselves at the onset of a phase transition. Here are the key points:
In Fig (3a), the minimum energy state has taken shape, but it hasn't yet become the global minimum.
The local minimum potential $\alpha$ corresponds to the small black hole.
But the environmental conditions evolve in such a way that a secondary minimum is formed.
Fig 3b shows intense movement of the small black hole toward the large black hole.
The transition probability ($\alpha\rightarrow\gamma$) dominates over the other transition.
We're witnessing the early stages, where the big black hole hasn't fully formed due to insufficient particles for reverse processes.
Fig 3c reveals that the structure is entirely under the influence of the transition from the small black hole to the large one.
The change in position ($\Delta r_k$) is entirely positive.
Then we examine a state midway through the phase transition (Fig 4a).
Although $\gamma$ not yet a global minimum, it rivals the alpha minimum in depth.
In a static, single-frame view, escape rates and movement probabilities from the local beta maximum (toward small or large black holes) appear equal.
However, environmental conditions and a shrinking radius favor the transition ($\alpha\rightarrow\gamma$) over the other (Fig 4b).
A significant point in this figure compared to the previous one, a collision point emerges between escape rate diagrams.
Beyond this point, the reverse process dominates in a small region at the diagram's end.
As we approach the phase transition's end, a negative region forms (Fig 4c), reinforcing feedback processes and altering initial conditions.
Near the phase transition's end, the system assumes a complete shape (Fig 5a) and $\gamma$ becomes a global minimum.
Environmental conditions favor the transition ($\beta\rightarrow\alpha$) (Fig 5b). In fact, the negative region (transition of a large black hole to a small black hole) has overcome the positive region (transition of a small black hole to a large black hole).
Remarkably, the contact point shifts significantly upward, emphasizing the reverse process (Fig 5c).
This quasi-oscillatory behavior prevents uncontrollable transitions and maintains black hole stability.
\section{AdS charged black hole with Barrow statistics}
In the realm of quantum gravity and black hole physics, Barrow's contributions are notable for incorporating quantum gravity effects on black hole structures. His work highlights how quantum gravity deforms the event horizon area of black holes\cite{36}. Barrow suggests that these effects cause the event horizon to adopt a fractal structure, challenging the traditional view of a smooth and simple event horizon.
This fractal deformation of the event horizon area led Barrow to revise the expression of black hole entropy. The deformation, characterized by a fractal parameter $\delta$, has been extensively discussed in both black hole and cosmological contexts. Notable discussions include constraints from the generalized second law of thermodynamics, Barrow holographic dark energy, and applications in the gravity-thermodynamics conjecture. Additionally, Barrow entropy has been applied to traditional black hole thermodynamics, thermodynamic geometries of BTZ black holes, charged AdS black holes with a global monopole, black hole quasinormal modes, and the equipartition theorem. According to the equations (5-8) that we mentioned, now we want to challenge the effect of Barrow entropy on the mentioned model and dynamic phase transition\cite{40,41,42,43}. After deriving the first law and the Smarr relation, we will reformulate the restricted phase space thermodynamics by incorporating Barrow’s corrections. Barrow has shown that quantum gravity effects can cause deformations on the black hole’s surface, described by a fractal structure. This leads to a modification in the black hole entropy expression, known as Barrow entropy\cite{44}.
\begin{equation}\label{(17)}
S_{B} = \big(\frac{A}{A_{pl}}\big)^{1+\frac{\delta}{2}}.
\end{equation}
In this context, \( A \) represents the event horizon area of a black hole, \( A_{PI} = 4G \) is the Planck area, and \( \delta \) quantifies the deformations due to quantum gravity on the event horizon area, with \( 0 \leq \delta \leq 1 \). When \( \delta = 0 \), the Bekenstein-Hawking entropy is restored, indicating no fractal structure. For \( \delta = 0 \), we return to the standard study of Reissner-Nordström AdS black holes, without a fractal structure on the event horizon area. Conversely, \( \delta = 1 \) signifies the most deformed and complex fractal structure of the black hole’s event horizon area. Due to the quantum gravity effects on the event horizon area, the black hole entropy becomes Barrow entropy. For the main quantities of this model i.e., entropy S, mass M, Hawking temperature T, and pressure P, we will have\cite{44}
\begin{equation}\label{(18)}
S_{B} = (\pi r_+^2)^{1+\frac{\delta}{2}}
\end{equation}

\begin{equation}\label{(19)}
M =\frac{8 r^{4} \pi  P +3 q^{2}+3 r^{2}}{6 r}
\end{equation}

\begin{equation}\label{(20)}
T =\frac{8 r^{4} \pi  P -q^{2}+r^{2}}{2 \pi  r^{3} \left(2+\delta \right) \left(\pi  r^{2}\right)^{\frac{\delta}{2}}}
\end{equation}
and,
\begin{equation}\label{(21)}
P =\frac{2 T \pi  r^{3} \left(\pi  r^{2}\right)^{\frac{\delta}{2}} \delta +4 T \pi  r^{3} \left(\pi  r^{2}\right)^{\frac{\delta}{2}}+q^{2}-r^{2}}{8 r^{4} \pi}
\end{equation}
Using equations (3) and (4) for the Gibbs and thermal potential of this model, we obtain:
\begin{equation}\label{(22)}
\widetilde{G} =\frac{8 r^{4} \pi  P +3 q^{2}+3 r^{2}}{6 r}-\frac{\left(8 r^{4} \pi  P -q^{2}+r^{2}\right) \left(\pi  r^{2}\right)^{1+\frac{\delta}{2}}}{2 \pi  r^{3} \left(2+\delta \right) \left(\pi  r^{2}\right)^{\frac{\delta}{2}}}
\end{equation}
and
\begin{equation}\label{(23)}
U =\frac{\pi^{\frac{\delta}{2}} \left(r^{2}\right)^{\frac{\delta}{2}} \left(8 r^{4} \pi  P +3 q^{2}+3 r^{2}\right)}{6 r \left(\pi  r^{2}\right)^{\frac{\delta}{2}}}-T \left(\pi  r^{2}\right)^{1+\frac{\delta}{2}}
\end{equation}
Critical quantities of the model which is given by,
\begin{equation}\label{(24)}
r_{c}=\frac{\sqrt{2}\, \sqrt{\left(1+\delta \right) \left(\delta +3\right)}\, q}{1+\delta}
\end{equation}

\begin{equation}\label{(25)}
P_{c}=-\frac{\left(1+\delta \right)^{2}}{32 q^{2} \left(\delta +3\right) \pi  \left(\delta -1\right)}
\end{equation}
and
\begin{equation}\label{(26)}
T_{c}=\frac{2^{\frac{1}{2}-\frac{\delta}{2}} \left(-1-\delta \right) \pi^{-1-\frac{\delta}{2}} \left(\frac{q^{2} \left(\delta +3\right)}{1+\delta}\right)^{-\frac{\delta}{2}}}{q \left(\delta +3\right) \sqrt{\left(1+\delta \right) \left(\delta +3\right)}\, \left(\delta^{2}+\delta -2\right)}
\end{equation}
So with respect to $q = 0.8, \delta = 0.5$, we can obtain,
$$r_c = 1.72819751912,\hspace{0.25cm}
P_c = 0.01998318202,\hspace{0.25cm}
T_c = 0.04810880076,\hspace{0.25cm}
U_c = 0.6912790083,\hspace{0.25cm}
G_c = 0.6912790081$$
Now, according to the above concepts, I will examine the role of Barrow entropy on dynamic phase transition.
To investigate the escape rate within this black hole, we consider a motion trajectory analogous to the previous state. We focus on two distinct scenarios to avoid redundancy: one occurring at the onset of the phase transition and the other near the conclusion of the phase transition. Our aim is to minimize repetitive descriptions. At the onset of the phase transition (Fig 8c), we observe a predominantly direct transition. The positive region dominates, and no reactive structures appear until the process concludes.  As the phase transition progresses (Fig 9). However, the negative region grows significantly, becoming comparable to the positive region in (Fig 9c). Similar to the previous case, reactive processes become more potent near the process's end.

\begin{figure}[h!]
 \begin{center}
 \subfigure[]{
 \includegraphics[height=6.5cm,width=8cm]{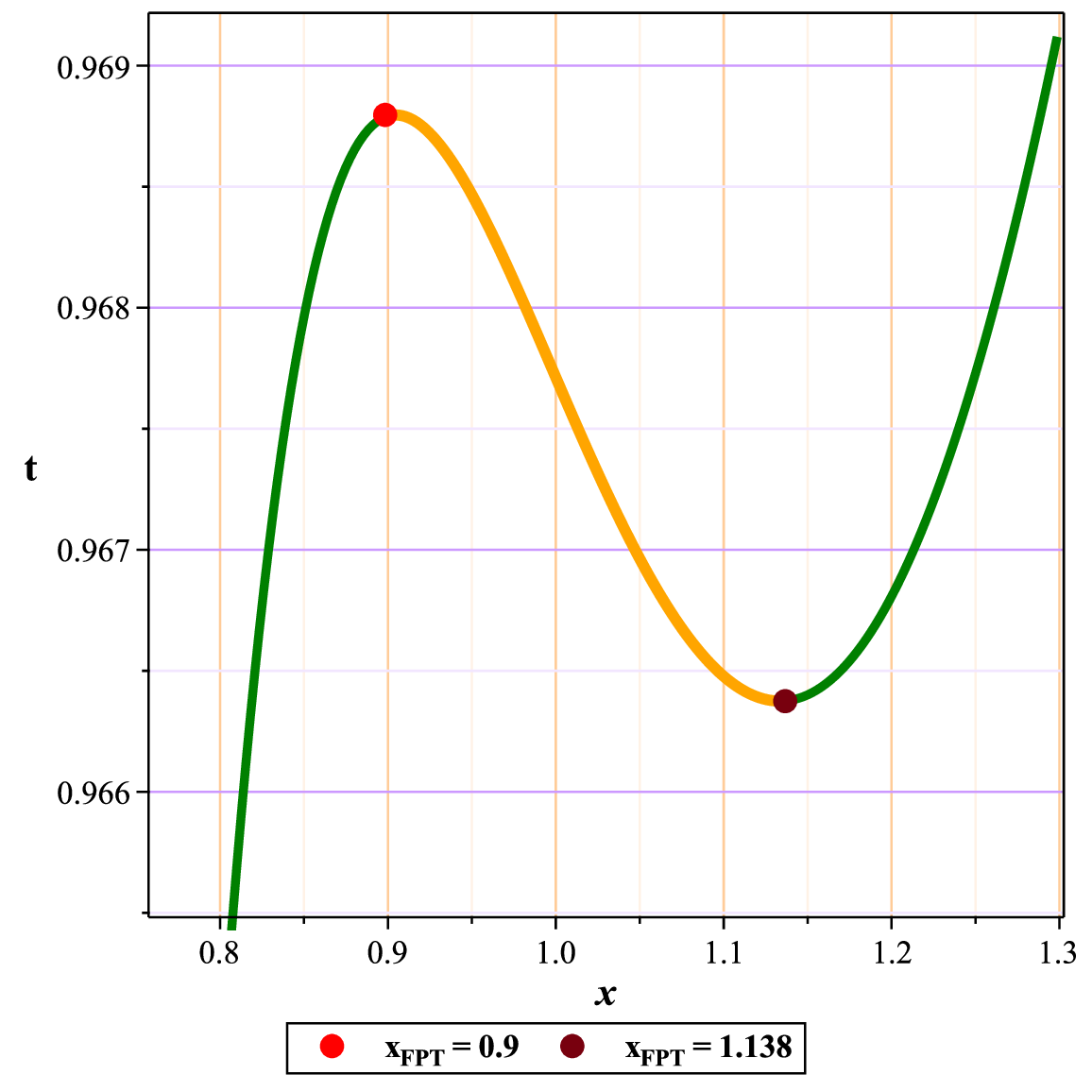}
 \label{fig6a}}
 \subfigure[]{
 \includegraphics[height=6.5cm,width=8cm]{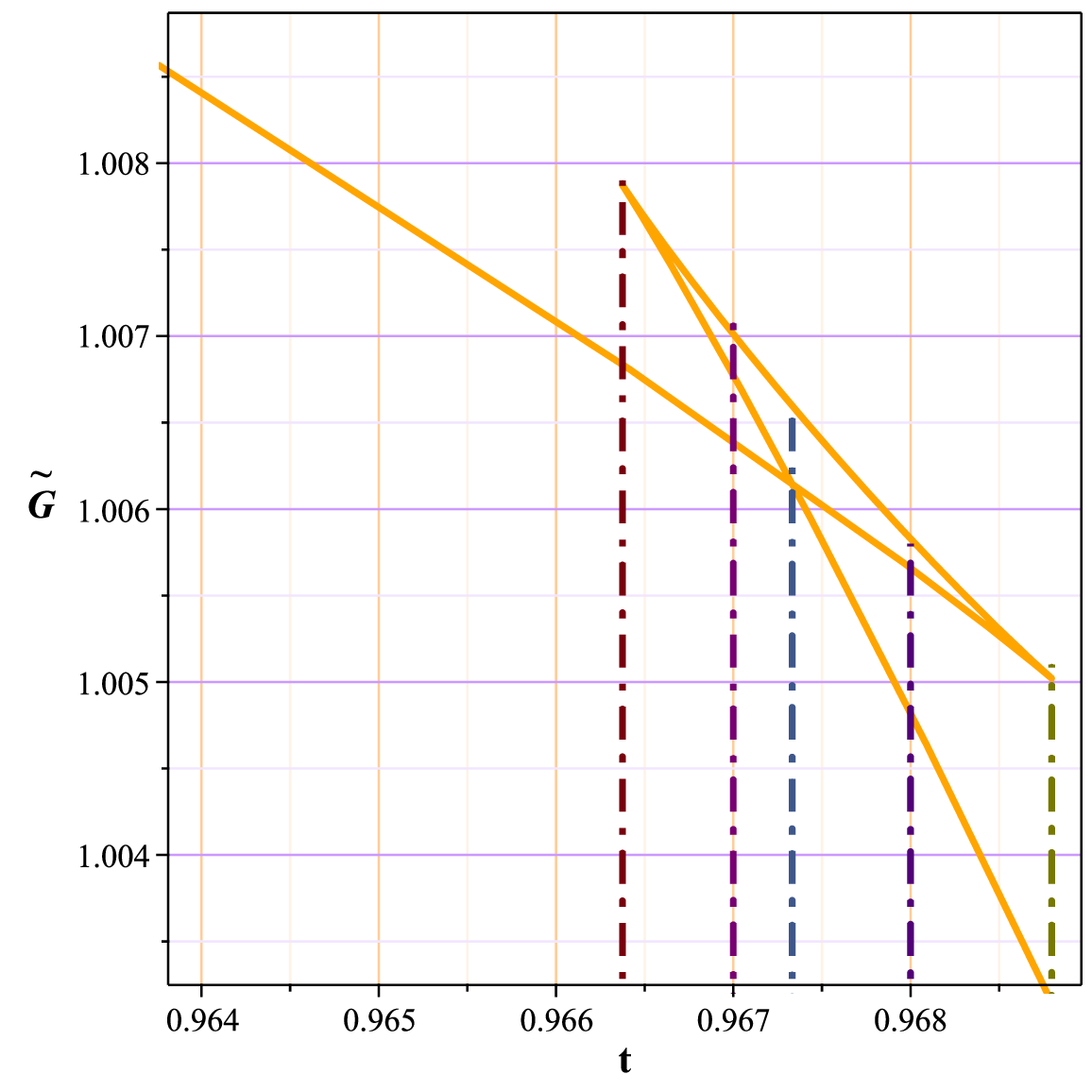}
 \label{fig6b}}
 \caption{\small{(6a): The plot of temperature ($T$) against the free parameters ($x$).
(6b): The representation of Gibbs free energy as a function of temperature.}}
 \label{fig6}
 \end{center}
 \end{figure}

\begin{figure}[h!]
 \begin{center}
 \subfigure[]{
 \includegraphics[height=6.5cm,width=6cm]{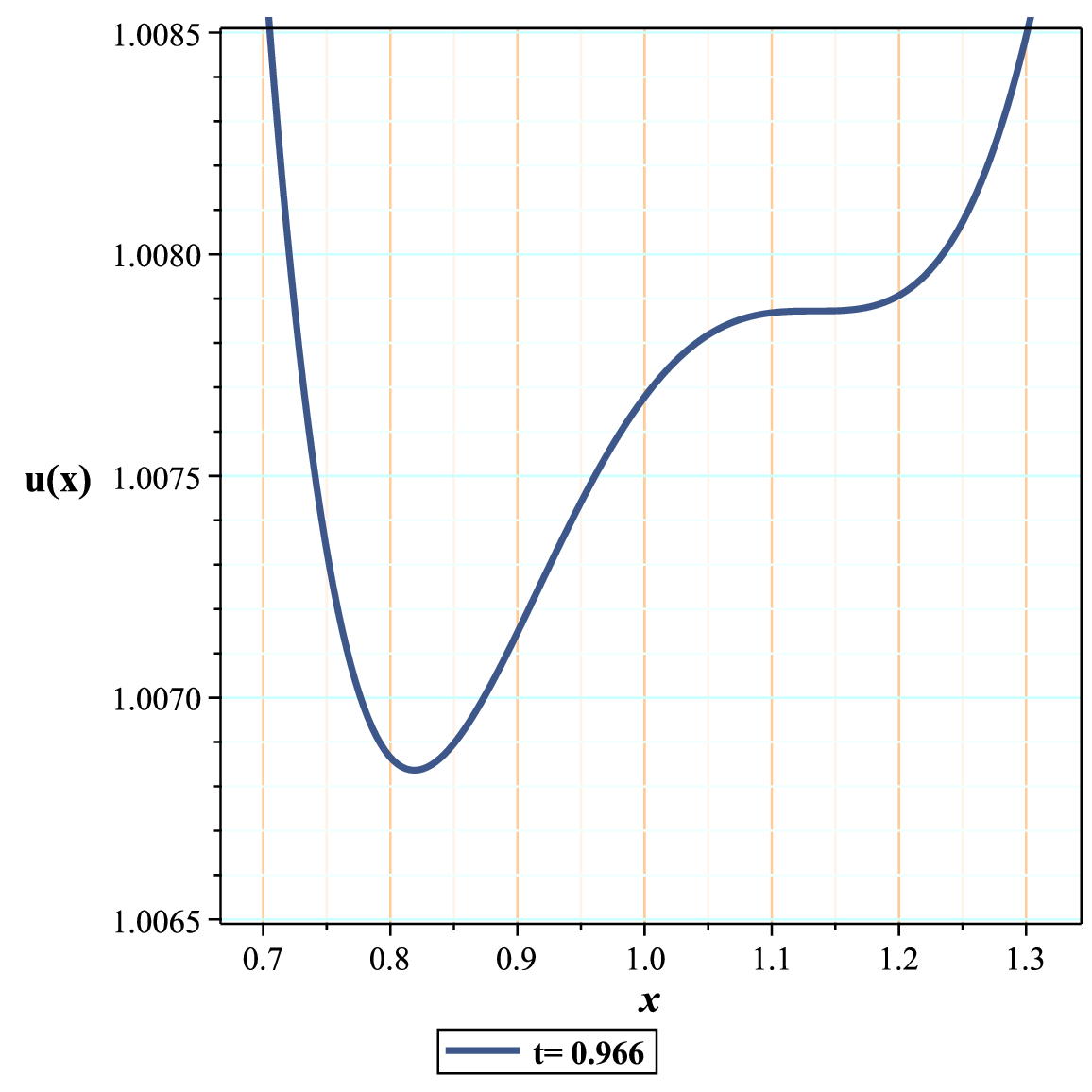}
 \label{fig7a}}
 \subfigure[]{
 \includegraphics[height=6.5cm,width=6cm]{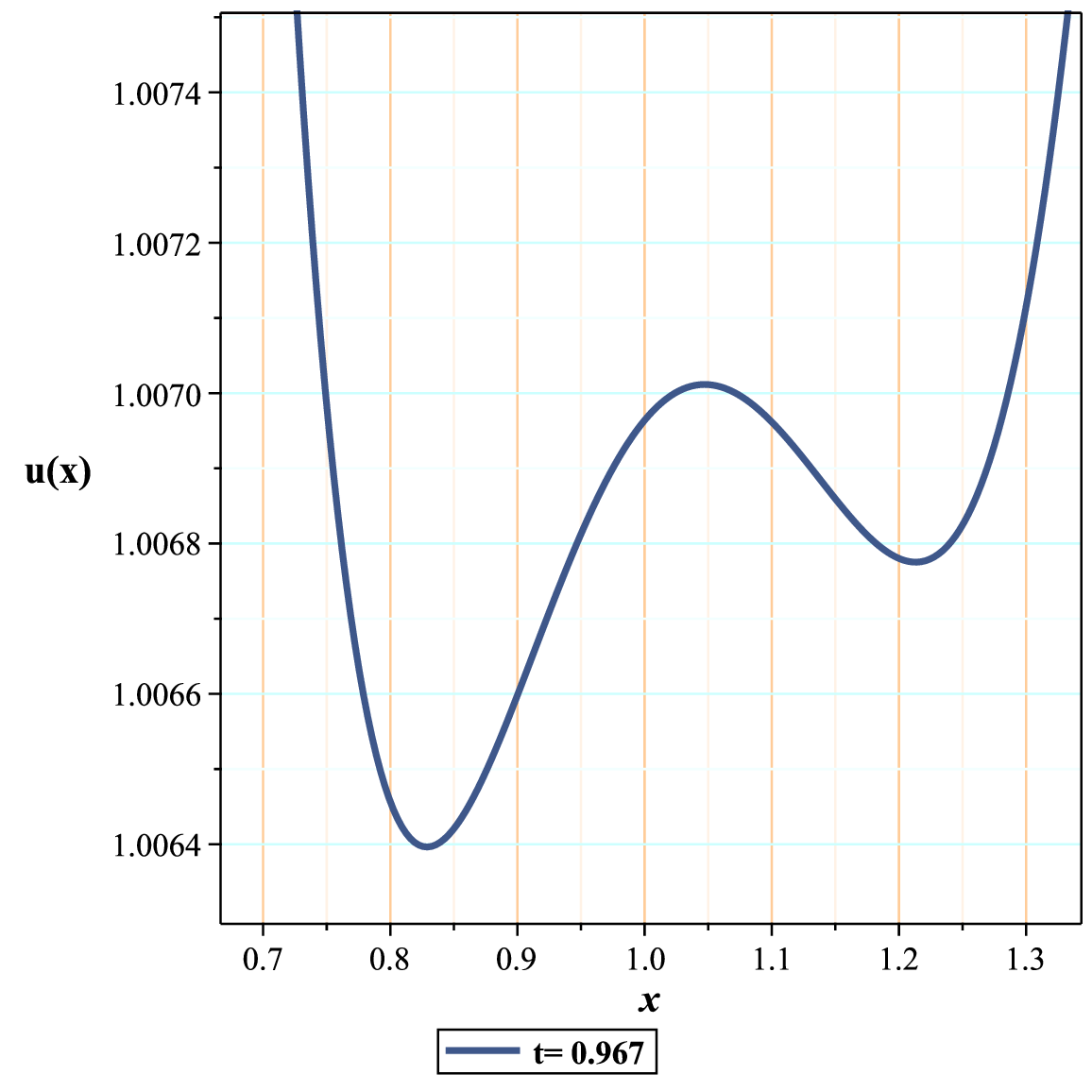}
 \label{fig7b}}
 \subfigure[]{
 \includegraphics[height=6.5cm,width=6cm]{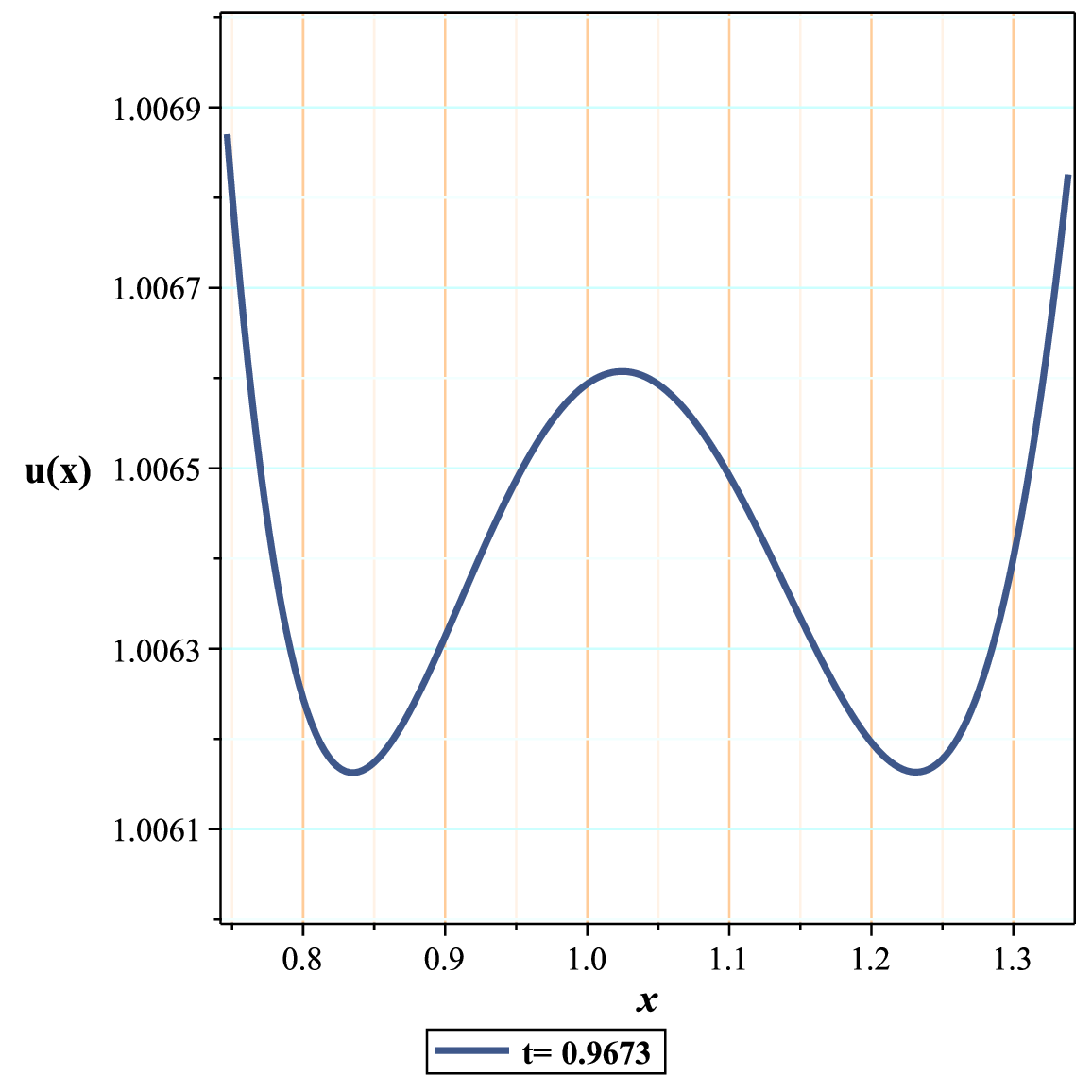}
 \label{fig7c}}
 \subfigure[]{
 \includegraphics[height=6.5cm,width=6cm]{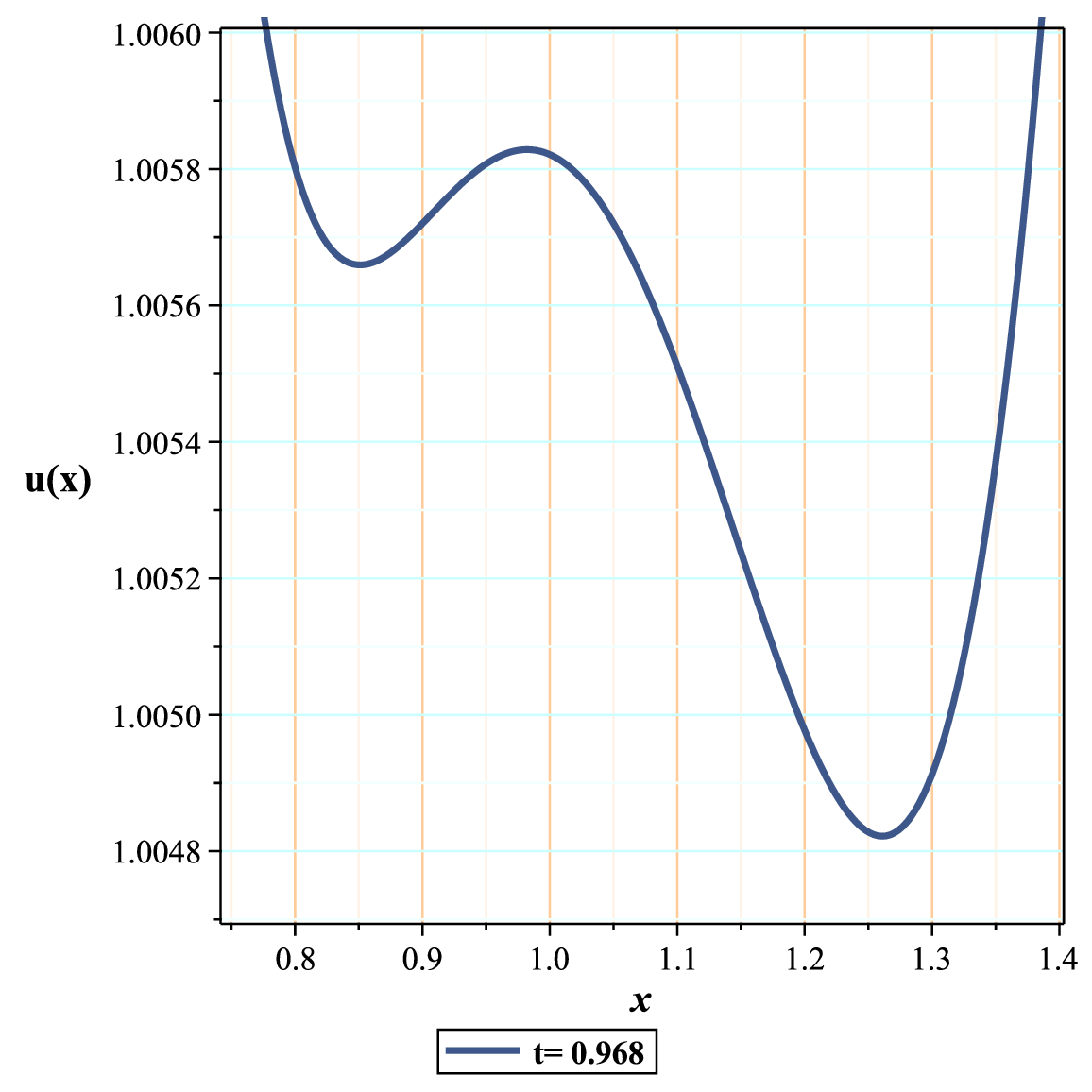}
 \label{fig7d}}
 \subfigure[]{
 \includegraphics[height=6.5cm,width=6cm]{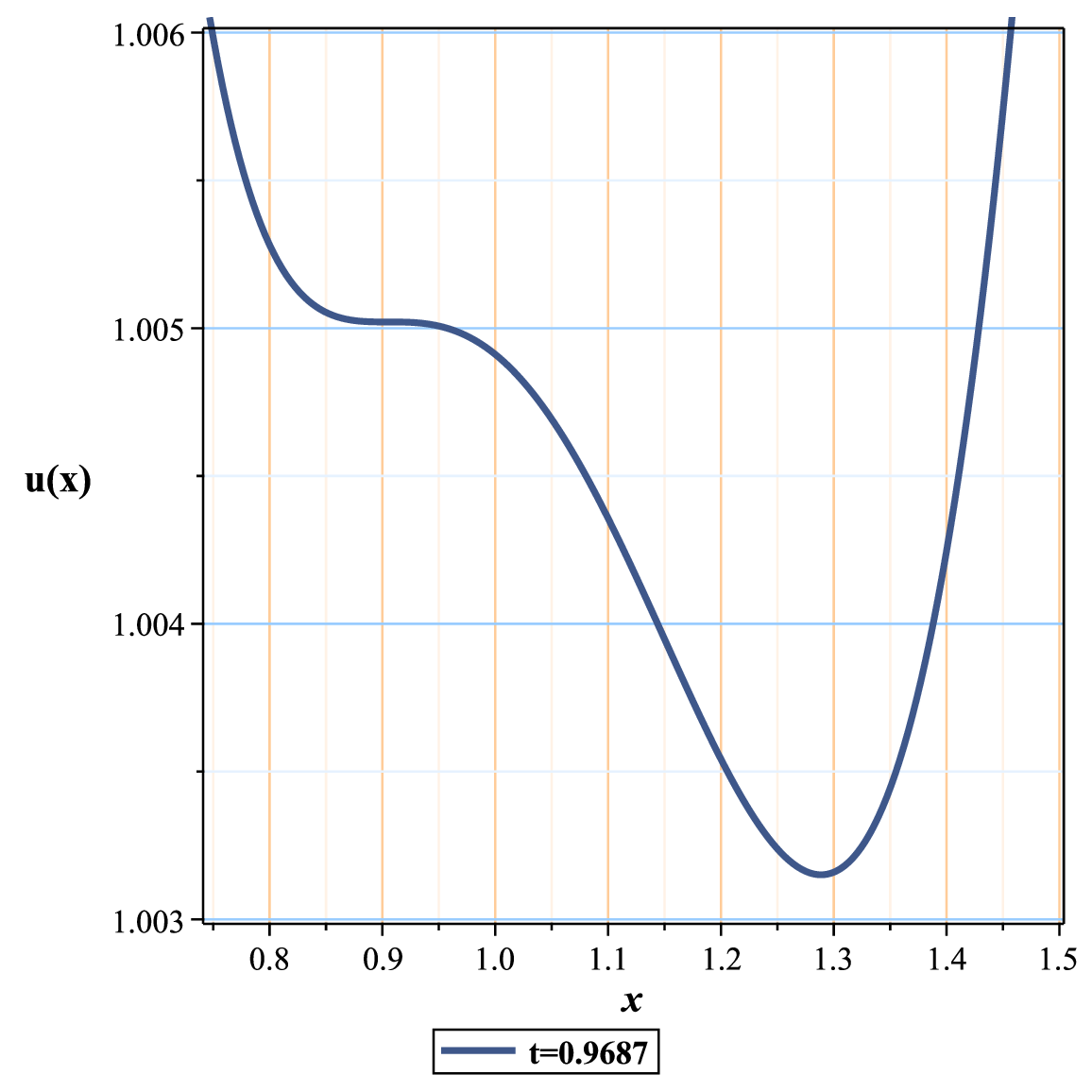}
 \label{fig7e}}
 \caption{\small{The variation of energy with respect to the parameter (x) at different temperatures}}
 \label{fig7}
 \end{center}
 \end{figure}

\begin{figure}[h!]
 \begin{center}
 \subfigure[]{
 \includegraphics[height=5cm,width=6cm]{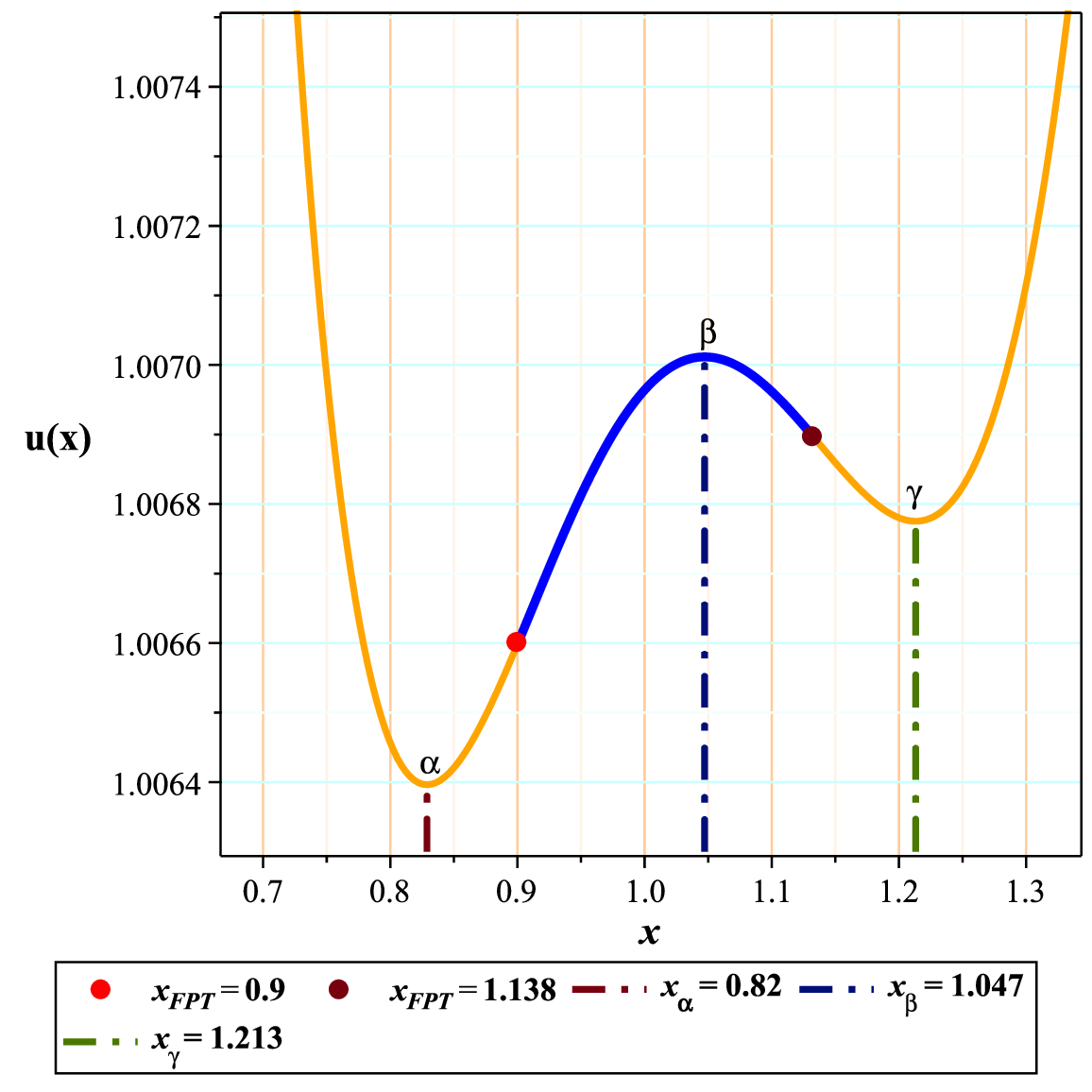}
 \label{fig8b}}
 \subfigure[]{
 \includegraphics[height=5cm,width=6cm]{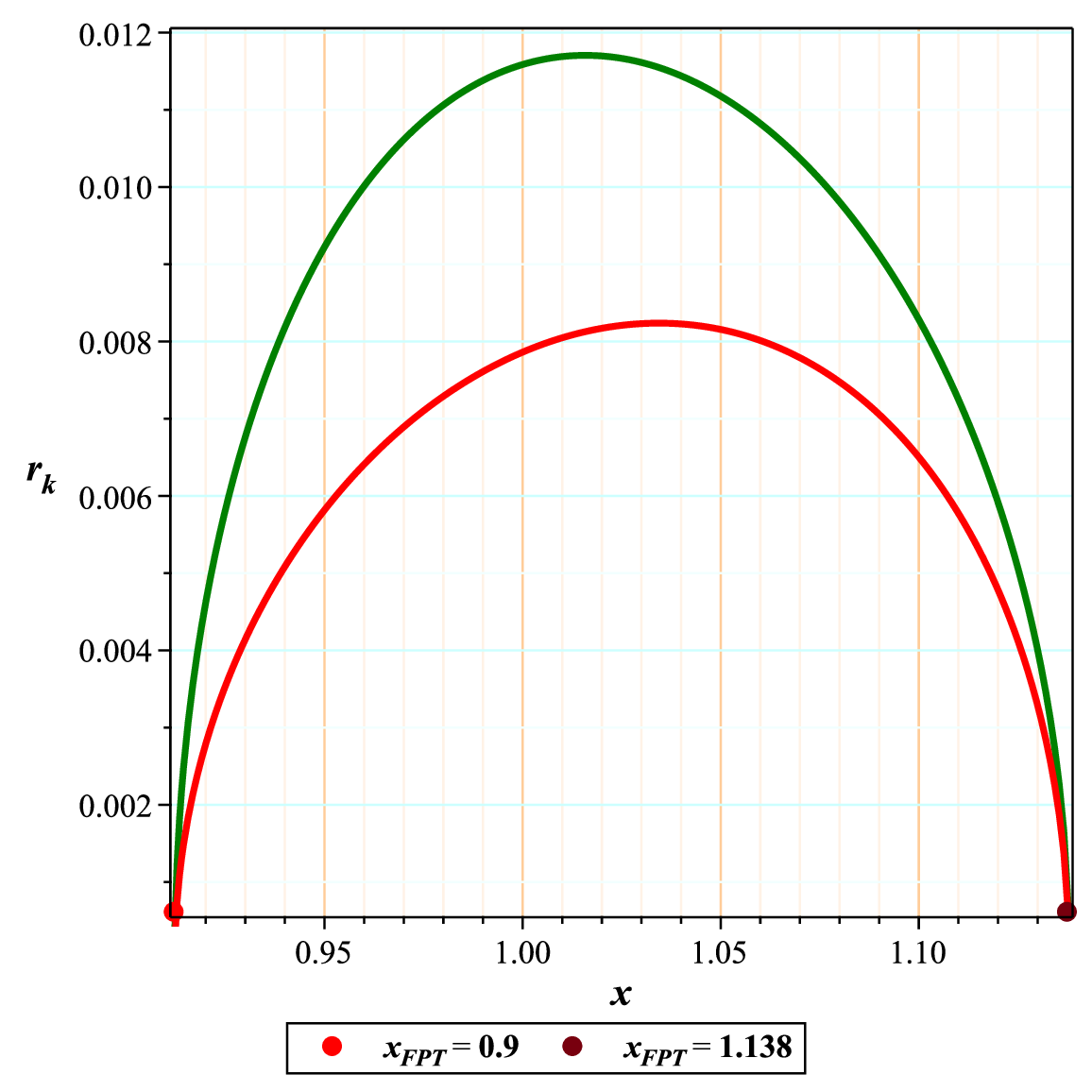}
 \label{fig8c}}
 \subfigure[]{
 \includegraphics[height=5cm,width=6cm]{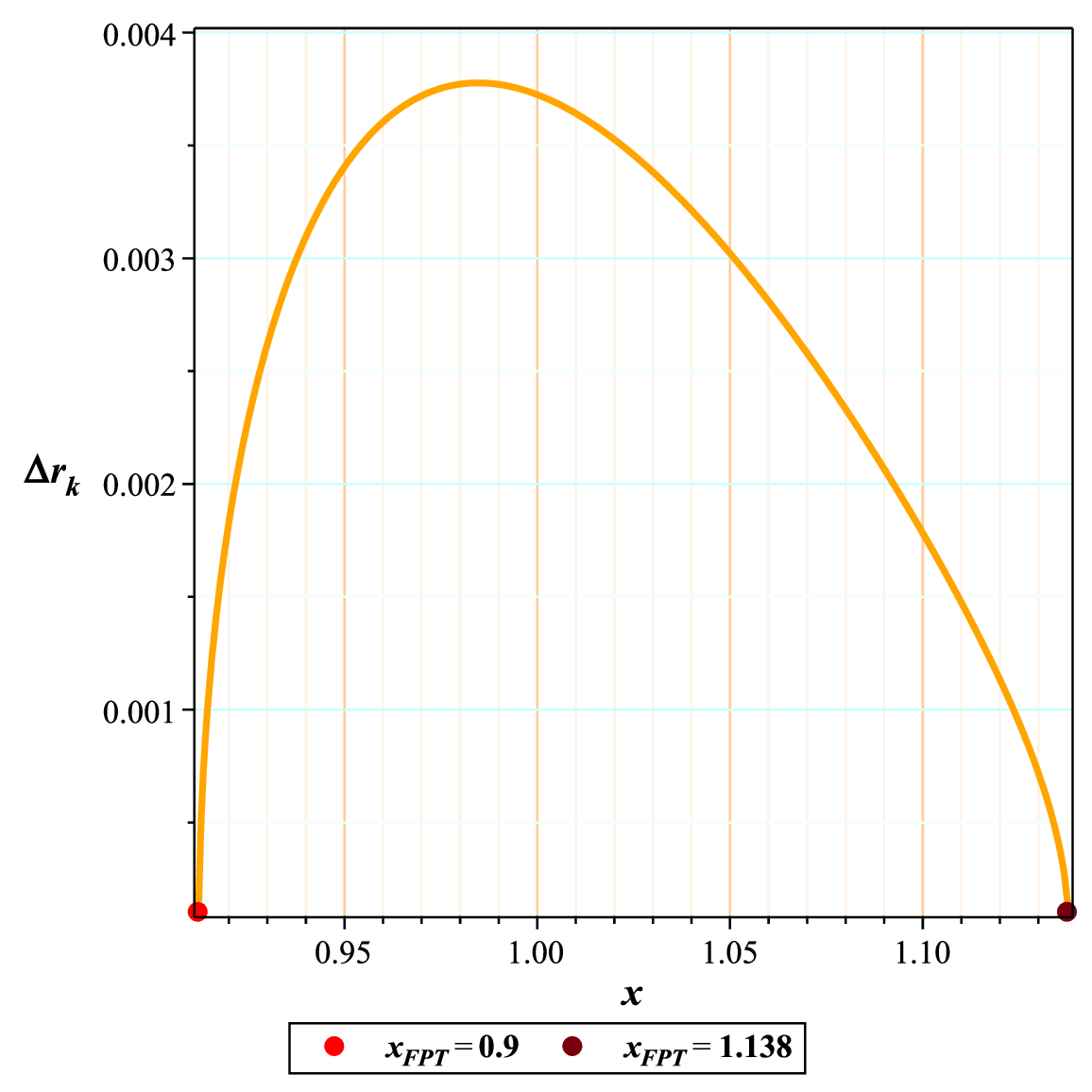}
 \label{fig8d}}
 \caption{\small{(8a): Illustrates the first-order phase transition region with respect to \(t\) and \(x\). It presents the plot of \(u(r)\) versus \(x\), considering free parameters and the First Passage Time (FPT) points.
(8b): Compares \(r_k\) as a function of \(x\) for two different conditions: \(\alpha \rightarrow \gamma\) and \(\gamma \rightarrow \alpha\). It also includes the coordinates of the contact point.
(8c): Depicts \(r_k\) as the difference between \(\Delta r_k\) for \(\alpha \rightarrow \gamma\) and \(\gamma \rightarrow \alpha\) as a function of \(x\).}}
 \label{fig8}
 \end{center}
 \end{figure}

\begin{figure}[h!]
 \begin{center}
 \subfigure[]{
 \includegraphics[height=5cm,width=6cm]{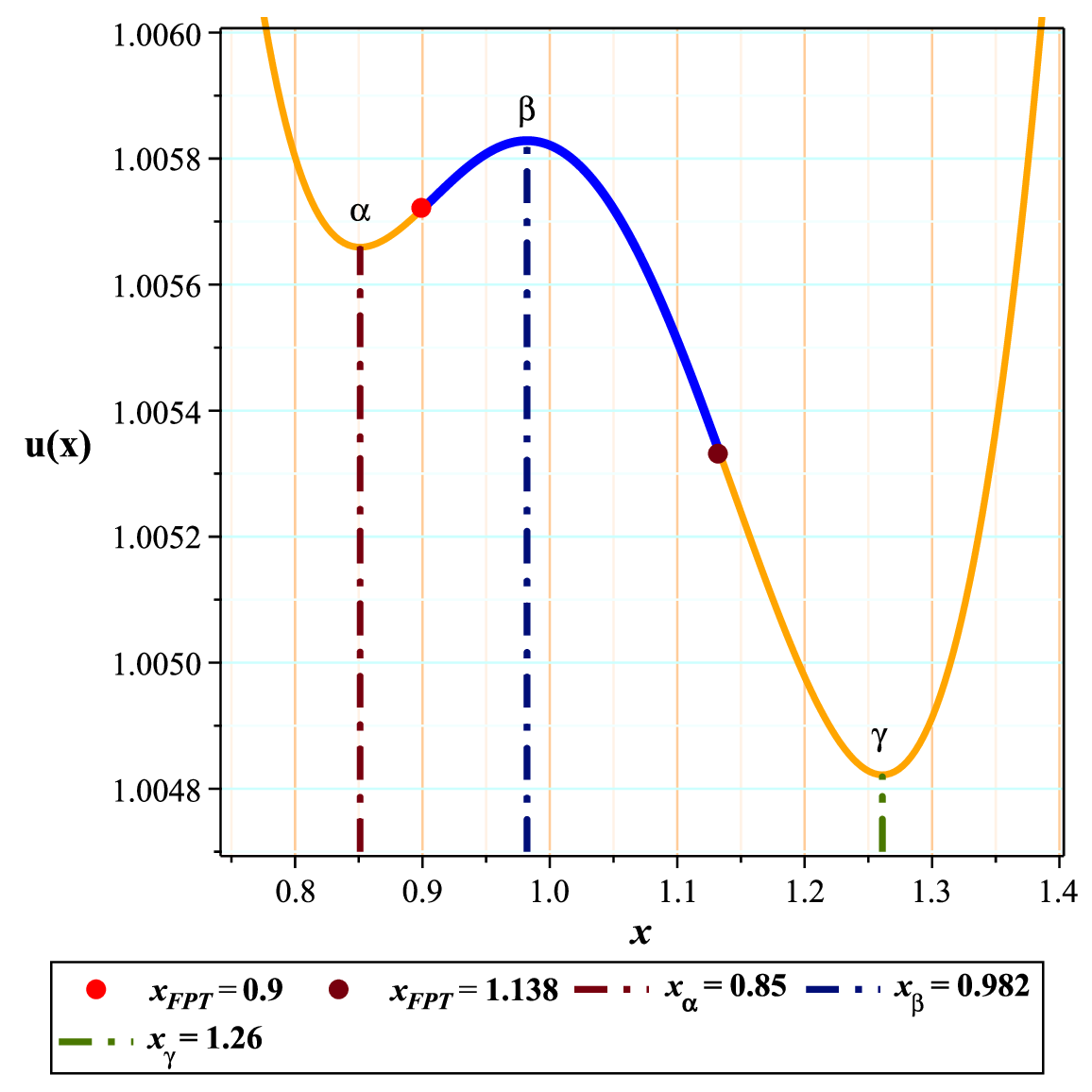}
 \label{fig9a}}
 \subfigure[]{
 \includegraphics[height=5cm,width=6cm]{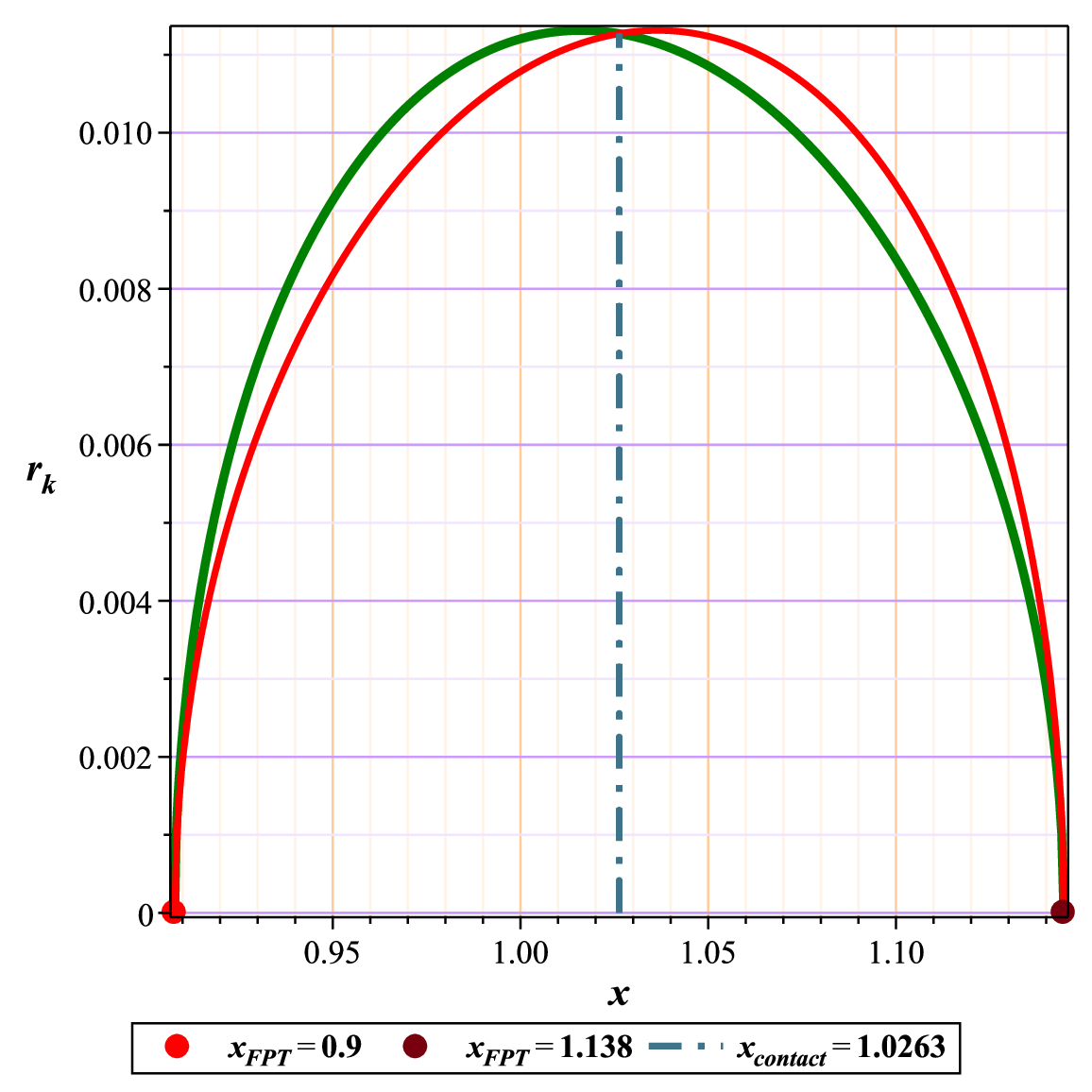}
 \label{fig9b}}
 \subfigure[]{
 \includegraphics[height=5cm,width=6cm]{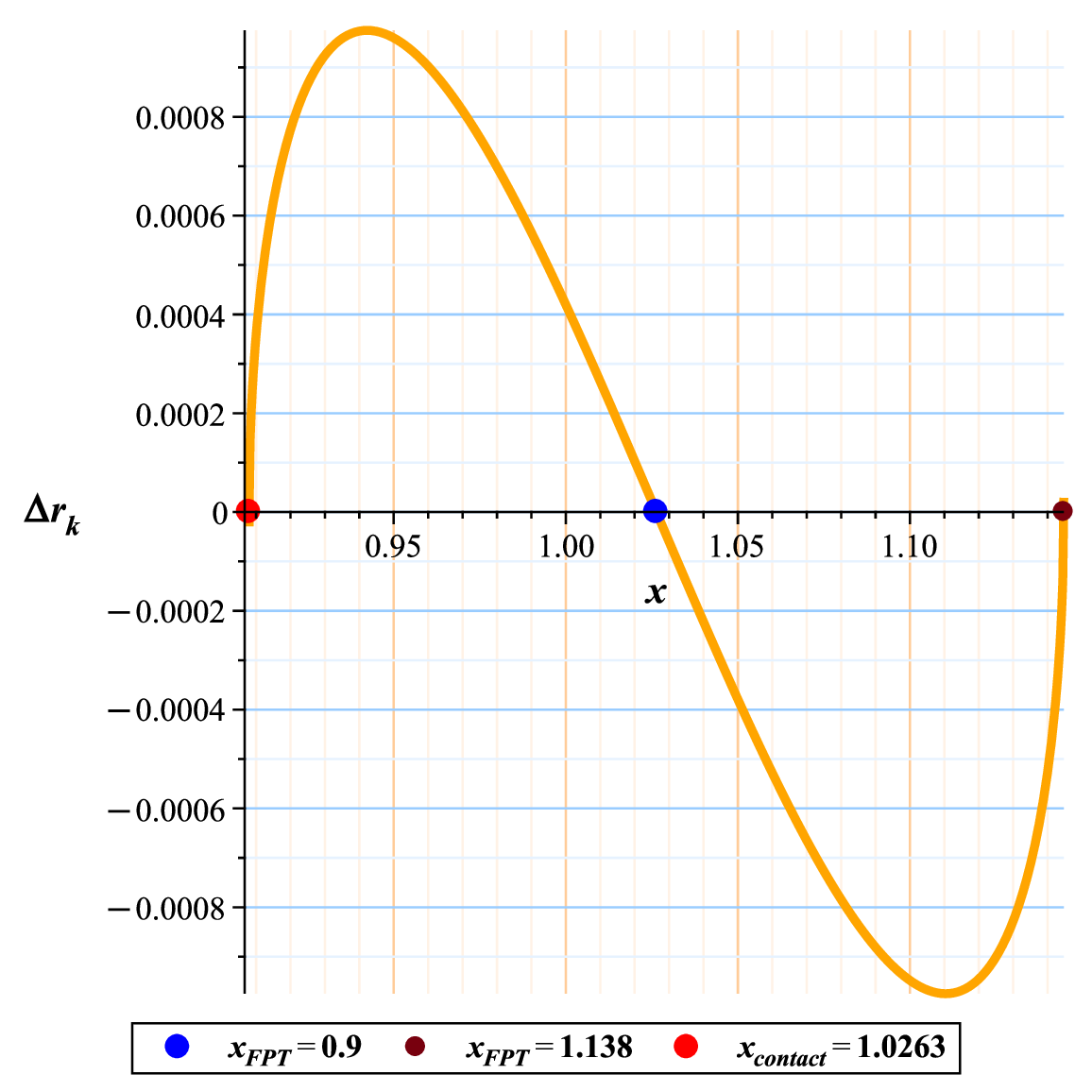}
 \label{fig9c}}
 \caption{\small{(9a): The plot of the function \(u(r)\) against \(x\) considering free parameters and the First Passage Time (FPT) points.
(9b): A comparison of \(r_k\) with respect to \(x\) for transitions from \(\alpha\) to \(\gamma\) and from \(\gamma\) to \(\alpha\).
(9c): The change in \(r_k\) (\(\Delta r_k\)) with respect to \(x\).}}
 \label{fig9}
 \end{center}
 \end{figure}
The results of the present work indicate that investigating dynamic phase transitions in AdS charged black holes, which are influenced by special structures such as Kaniadakis and Barrow statistics, yields results very close to those of the conventional model in liturature\cite{23}. Consequently, these structures can be further studied to explore other cosmological properties and different types of phase transitions.
\section{Conclusions}
In this study, we delve into the fascinating realm of black hole thermodynamics, specifically focusing on phase transitions. Our analytical lens centers on anti-de Sitter (AdS) charged black holes, which exhibit intriguing behavior when influenced by Kaniadakis and Barrow statistics. Let's explore the key findings:\\
1) The Kramers Escape Rate Method: We employ the Kramers escape rate method—a powerful tool for understanding phase transitions in complex systems. By tracking the transition probabilities between different states, we gain insights into the underlying dynamics. The Kramers escape rate is typically applied to particles experiencing escape from a potential well due to thermal fluctuations. However, in this  and previous study\cite{23}, we observe that this rate aligns well with the first-order phase transition of a black holes. This phase transition may occur without particles or even continuously. This harmony could provide significant motivation for extending the use of this rate to other potential-like structures in gravity and cosmology that exhibit similar behavior.\\ 2) Dominance of Large Black Hole Formation: Notably, the transition from small to large black holes dominates the phase transition process. This transition is pivotal in shaping the thermodynamic landscape of AdS black holes.\\ 3) Addressing Stochastic Process Gaps: Our research bridges gaps in stochastic process analysis related to phase transition rates. By considering Kaniadakis and Barrow statistics, we contribute to a deeper understanding of the intricate interplay between statistical mechanics and black hole thermodynamics.\\ 4) Applications of Kaniadakis and Barrow statistics. We propose potential applications in cosmological contexts, where these structures may shed light on other fundamental properties of the universe.
Surprisingly, our results reveal that dynamic phase transitions in AdS-charged black holes, influenced by these special statistical structures, yield outcomes remarkably close to those predicted by conventional models\cite{23}. This alignment underscores the significance of exploring alternative statistical frameworks. In summary, our investigation not only enriches our understanding of black hole thermodynamics but also opens avenues for further exploration. By scrutinizing these unique structures, we unlock new insights into cosmological phenomena and diverse phase transitions.

\end{document}